\documentclass[acmsmall]{acmart}

\AtBeginDocument{%
  \providecommand\BibTeX{{%
    \normalfont B\kern-0.5em{\scshape i\kern-0.25em b}\kern-0.8em\TeX}}}

\setcopyright{acmcopyright}
\copyrightyear{2023}
\acmYear{2023}
\acmDOI{XXXXXXX.XXXXXXX}

\usepackage[printonlyused,nolist,nohyperlinks]{acronym}
\begin{document}
\acrodef{ML}{machine learning}
\acrodef{FMEA}{Failure Mode and Effects Analysis}
\acrodef{STPA}{System Theoretic Process Analysis}
\title[From plane crashes to algorithmic harm]{From plane crashes to algorithmic harm: applicability of safety engineering frameworks for responsible ML}
\author{Shalaleh Rismani}
\affiliation{%
 \institution{Google Research, McGill University}
 \city{Montreal}
 \country{Canada}
}

\author{Renee Shelby}
\affiliation{%
  \institution{Google, JusTech Lab, Australian National University}
  \city{San Francisco}
  \country{U.S.A}
}

\author{Andrew Smart}
\affiliation{%
  \institution{Google Research}
  \city{San Francisco}
  \country{U.S.A}
}

\author{Edgar Jatho}
\affiliation{%
  \institution{Naval Postgraduate School}
  \city{Monterey}
  \country{U.S.A}
}

\author{Josh A. Kroll}
\affiliation{%
  \institution{Naval Postgraduate School}
  \city{Monterey}
  \country{U.S.A}
}

\author{AJung Moon}
\affiliation{%
  \institution{McGill University}
  \city{Montreal}
  \country{Canada}
}

\author{Negar Rostamzadeh}
\affiliation{%
   \institution{Google Research}
   \city{Montreal}
   \country{Canada}
}
\renewcommand{\shortauthors}{Rismani et al.}

\begin{abstract}
Inappropriate design and deployment of machine learning (ML) systems leads to negative downstream social and ethical impact -- described here as social and ethical risks -- for users, society and the environment. 
Despite the growing need to regulate ML systems, current processes for assessing and mitigating risks are disjointed and inconsistent. 
We interviewed 30 industry practitioners on their current social and ethical risk management practices, and collected their first reactions on adapting safety engineering frameworks into their practice -- namely, System Theoretic Process Analysis (STPA) and Failure Mode and Effects Analysis (FMEA). 
Our findings suggest STPA/FMEA can provide appropriate structure toward social and ethical risk assessment and mitigation processes. However, we also find nontrivial challenges in integrating such frameworks in the fast-paced culture of the ML industry. We call on the ML research community to strengthen existing frameworks and assess their efficacy, ensuring that ML systems are safer for all people.
\end{abstract}



\maketitle

\section{Introduction}
During a panel at the 1994 ACM Conference on Human Factors in Computing Systems (CHI), prominent scholars from different disciplines convened to discuss "what makes a good computer system good." Panelists highlighted considerations for safety, ethics, user perspectives, and societal structures as critical elements for making a \textit{good} system \cite{Friedman1994-rm}. Almost 28 years later, we posit that these epistemological perspectives need to be in deeper conversation for designing and assessing \ac{ML} systems that challenge conventional understanding of safety and harm. 

The development and use of ML systems can adversely impact people, communities, and society at large ~\cite{Ehsan2022-gx,Weidinger2022-qv,Noble2018-xs,Mohamed2020-uu,Suresh2021-js,Blodgett2022-rc}, including inequitable resource allocation ~\cite{Angwin2016-td,Sambasivan2021-gm,Chen2016-tb}, perpetuating normative narratives about people and social groups ~\cite{Kay2015-wn,Tatman2017-eg}, and the entrenchment of social inequalities ~\cite{Abid2021-ls,Mann2019-ns,Liang2020-xf}. We frame these adverse impacts broadly as \emph{social and ethical risks}. To manage such risks, quantitative ~\cite{Franklin2022-qe, Lee2021-oo}, qualitative  ~\cite{Liao2020-cy,Gebru2021-kx,Mitchell2019-ms,Raji2020-dw}, and epistemological frameworks ~\cite{Ehsan2021-st, Mohamed2020-uu,Hanna2020-rn} have been proposed. In parallel, active regulatory and standards activities are taking place internationally ~\cite{Government_of_Canada2022-bq, European_Commission2021-mv, NIST,ieee, ISO}. Despite the rapidly evolving discourse, there is limited empirical understanding of how proposed social and ethical risk management tools have been adopted by practitioners ~\cite{Madaio2022-yz,Feder_Cooper2022-fh}. 

Inspired by the 1994 panelists, we examine the dialogue between safety engineering frameworks and understandings of social and ethical risks of \ac{ML} systems. 
First, we report on ethical and social risk management practices currently used in the industry. Second, we take a developmental approach to examine how safety engineering frameworks can improve existing practices. We chose two of the most successful safety engineering frameworks used in other sociotechnical domains ~\cite{Sulaman2019-rm,Pawlicki2016-ef,Bugalia2022-hr}: Failure Mode and Effect Analysis (FMEA) ~\cite{Carlson2012-yh} and System Theoretic Process Analysis (STPA) ~\cite{Leveson2018-no,Patriarca2022-mv}, which we describe in detail in Section \ref{background}. 

We conducted 30 semi-structured in-depth interviews with industry practitioners who shared their current practices used to assess and mitigate social and ethical risks. We introduced the two safety engineering frameworks, inviting them to envision how they might employ them to assess ethical and social risk of ML systems. 
The results of our study address the following research questions:

\begin{itemize}
    \item \textbf{RQ1}: Which practices do ML practitioners use to manage social and ethical risks today?
    \item \textbf{RQ2}: What challenges do practitioners face in their attempts to manage social and ethical risks? 
    \item \textbf{RQ3}: How could safety engineering frameworks such as FMEA and STPA inform and improve current practices? What advantages and disadvantages of each method do ML practitioners identify? 
\end{itemize}
We contribute to the emerging research on managing social and ethical risk of \ac{ML} systems in human-computing scholarship and responsible ML communities by offering:

\begin{itemize}
    \item An overview of how practitioners define, assess and mitigate social and ethical risks;
    \item An analysis of the corresponding challenges when implementing these practices; 
    \item A set of insights on how FMEA and STPA could inform existing practices along with their perceived advantages and disadvantages;
    \item Future research directions and calls to action for HCI and responsible ML scholars. 
\end{itemize}

Our findings illustrate safety engineering frameworks provide valuable structure for investigating how social and ethical risks emerge from ML systems design and integration in a given context. However, successful adaptation of these frameworks requires solutions to existing organizational challenges for operationalizing formal risk management practices. Moreover, results of our work motivate further theoretical and applied research on adaptation of such frameworks. The remainder of this paper is organized as follows. We start by providing an overview of current discourse in responsible ML development and contextualize the relevance of the safety engineering frameworks (Section \ref{background}). We outline our interview protocol and analysis methods in Section \ref{methodology}, followed by highlighting key findings in Section \ref{findings}. We discuss the value and shortcomings of applying safety engineering frameworks in light of current practices and call on the research community to further examine and strengthen these frameworks for ethical and social risk management of \ac{ML} systems in Section \ref{discussion}. 

\section{Background}
\label{background}

Analyzing social and ethical implications of algorithmic systems is not new to computing researchers and practitioners  ~\cite{Friedman1996-ou,Barocas2013-fe,DIgnazio2021-ls,Orlikowski2000-ic}. 
In the literature, terms such as harm ~\cite{Suresh2021-js}, failure ~\cite{Raji2022-cz}, and risk ~\cite{Weidinger2022-qv,NIST} are often used to describe adverse impacts of ML systems. While there is currently no agreed upon definition of these terms and their relationships, we use the phrase \textit{social and ethical risk} to frame broadly the adverse social and ethical implications ML systems can have on users, society, and the environment. This working definition provides conceptual consistency in this paper, and is not meant to be normative. In the remainder of this section, we contextualize current discourses on social and ethical risks in ML to situate our study design, findings, and discussion. We highlight current epistemological perspectives and tools for responsible \ac{ML} development, and detail the safety engineering frameworks (FMEA and STPA).

\subsection{Epistemological perspectives for anticipating and mitigating harms of ML systems}
Scholars have proposed various methods for anticipating social and ethical impacts  ~\cite{Selbst2019-dr,Floridi2020-gu,Feder_Cooper2022-fh}. Anticipating harm involves thinking about the values ~\cite{Shilton2013-ez,Nissenbaum2001-dm} and affordances of ML systems ~\cite{Brey2012-pp}, with specific attention to how social norms and power dynamics constitutively shape adverse impacts of ML systems ~\cite{Benjamin2019-qt,Birhane2021-kh}. The process of anticipation is aided by critical epistemologies that center the needs and standpoints of socially oppressed groups, including critical race theory ~\cite{Hanna2020-rn,Benjamin2019-qt,Noble2018-xs,Klumbyte2022-tf}, post-colonial theories ~\cite{Mohamed2020-uu}, and queer ~\cite{Spiel2019-jk} and feminist HCI ~\cite{Bardzell2010-ky}. 

As social and ethical impacts are co-constituted through the interplay of technical system components and the social world ~\cite{Jasanoff2004-fy}, design methodologies attentive to these dynamics support more meaningful harms anticipation and mitigation. For instance, Value Sensitive Design that examines what value tensions ML systems create or resolve ~\cite{Friedman2019-jw, Wambsganss2021}, supports increased stakeholder coordination ~\cite{Umbrello2019-vn} and consideration of technology from different social standpoints and perspectives \cite{Ballard2019}. Similarly, participatory design methods can center the needs of users, communities, and other stakeholders often excluded from the design process \cite{Zytko2022} or algorithmic governance ~\cite{Lee2019procedural, Lee2019webuild}, especially when incorporating feminist epistemologies ~\cite{Bardzell2010-ky, Hope2019}. Speculative design can also help designers imagine more socially just and racially equitable technological futures ~\cite{Harrington2022}. While these epistemological perspectives and frameworks do not explicitly assess risk, they provide theoretical grounds for examining and mitigating social and ethical risk. 

\subsection{Responsible ML tools, processes, and emerging regulations}
With increased deployment of ML systems and reported harms \cite{Noble2018-xs,Weidinger2022-qv,Blodgett2022-rc}, there is movement towards formalizing quantitative and qualitative tools for responsible \ac{ML} development. 
Traditionally, \ac{ML} system evaluations ~\cite{hutchinson2022evaluation,rostamzadeh2021thinking} prioritized assessing and optimizing for a narrow set of performance metrics, mistakenly treating these measurements (e.g., accuracy of a test set) as a target rather than proxy for certain risks ~\cite{liao2021we}. Recognizing these shortcomings ~\cite{Jobin2019-qf}, \ac{ML} scholars proposed alternative methods to enable more comprehensive evaluation. These methods include assessing computational fairness with alternative statistical definitions ~\cite{corbett2017algorithmic, card2020consequentialism, chohlas2021learning, corbett2018measure}, quantifying model interpretability based on statistical properties ~\cite{Molnar2020-rb,Poursabzi-Sangdeh2018-ks}, evaluating robustness to distribution shift ~\cite{sugiyama2007direct,koh2020wilds,chen2021mandoline} and examining model performance when exposed to adversarial examples ~\cite{ruiz2022simulated, zeng2021openattack, zhang2020adversarial, ettinger2017towards}. 

In parallel, significant effort has also focused on developing mixed-method (qualitative and quantitative) processes to increase accountability and assess ML systems contextually. Scholars have proposed model cards ~\cite{Mitchell2019-ms}, datasheets ~\cite{Gebru2021-kx} and auditing tools ~\cite{Raji2020-dw,Shen2021-us,Brown2021-lo} to improve the transparency and quality of model and data practices. Human right and algorithmic impact assessments aid identification of potential societal level harms by examining model deployment in a given context ~\cite{Mantelero2022-en,Latonero2021-fl,Moss2021-qi}. Similarly, scholars have developed contextual methods of assessing fairness by focusing attention on the situated power dynamics of where systems are deployed ~\cite{sambasivan2021re,Yee2021-eg} and transparency ~\cite{Vaughan2020-zs} of \ac{ML} systems. Parallel to technique development, there is a rapidly emerging set of international standards \cite{ISO}, policies ~\cite{Treasury_Board_of_Canada_Secretariat2019-sk}, and regulatory frameworks ~\cite{European_Commission2021-mv} that examine ML systems from a risk-based perspective. 

\subsubsection{Empirical studies of responsible ML practices}
HCI scholarship examining the perceptions and needs of responsible ML practitioners have identified key challenges ~\cite{Rakova2021, Holstein2019}, including limited definitional consensus on key terms \cite{Krafft2020} and the underlying need to translate principles into actionable guidance to catalyze transformative organizational change ~\cite{Madaio2020, Kramer2018}. Practitioners often work in multidisciplinary environments, where technical and non-technical stakeholders draw on different epistemologies and perspectives ~\cite{Nahar2022}, posing challenges to cohesive anticipation and identification of harms and risks ~\cite{Wong2022}. In terms of risk assessment specifically, Raji et al. ~\cite{Raji2020-dw} underscore how the often-rapid pace and piecemeal implementation of risk assessment inhibits holistic forecasting of potential risks and their relationships to technical system components.

While there is a growing literature on practitioner needs, limited work has focused on identifying existing social and ethical risk management practices and ML practitioners’ perspectives towards safety engineering frameworks. Martelaro et al.’s ~\cite{Martelaro2022} study of the applicability of hazard analysis and the needs of practitioners is a notable exception. From an exploratory interview study with eight participants, Martelaro et al. conclude existing hazard analysis tools from safety engineering cannot readily support ML systems and highlight how lack of team incentives, the pace of industry development, and underestimating the effort needed to create robust ML systems challenge implementation of these tools. Nonetheless, Martelaro et al. emphasize frameworks are necessary to support risk management for responsible ML practice.
 
\subsection{Introducing safety engineering approaches to failure and hazard analysis}
Safety engineering is a generic term for an assemblage of engineering analyses and management practices designed to control dangerous situations arising in sociotechnical systems ~\cite{bahr2014system, ericson2015hazard, leveson2016engineering}. These analyses and practices identify potential hazards or system failures, understand their impact on users or the public, investigate causes, develop appropriate controls to mitigate the potential harms, and monitor systems ~\cite{shrader1991risk}. Safety engineering crystallized as a discipline around WWII, when military operators recognized losses and accidents were often the result of avoidable design flaws in technology and human factors ~\cite{vaughan1996challenger}. Since then, implementation of safety engineering in sociotechnical domains, such as medical devices and aerospace, has significantly reduced accidents and failures ~\cite{rodrigues2012commercial}. 

We motivate use of safety engineering for social and ethical risk management given its strength in drawing attention to the relationships between risks, system design, and deployment ~\cite{Dobbe2022-kf, Raji2020-dw}. As ML systems introduce interdependencies between the ML artifact, its operational environments, and society at large ~\cite{reader2022models}, safety frameworks can provide a strong analytical grounding for risk management ~\cite{ericson2015hazard}. Moreover, harms from ML systems are often recognized after they have occurred ~\cite{Raji2020closing} at which point mitigating them is significantly more challenging and costly ~\cite{Carlson2012-yh}. In this study, we focus on two safety engineering techniques designed to identify and address undesired outcomes early in development ~\cite{bahr2014system, ericson2015hazard, leveson2016engineering}: a failure analysis technique for improving reliability (FMEA) and a hazard analysis technique for identifying unsafe system states (STPA).

\subsubsection{Failure Mode and Effects Analysis (FMEA)}
FMEA, a long-standing reliability framework, takes an analytic reduction (i.e. divide and conquer) approach to identifying and evaluating likelihood of risk for potential failure modes (i.e. the mechanism of failure) for a technological system or process ~\cite{Carlson2012-yh}. FMEA has been used in high consequence projects, such as  space shuttle ~\cite{Jenab2015-ts} and U.S. nuclear power plant safety ~\cite{Hindawi_undated-gp}. The FMEA framework helps uncover potential failure modes, identify the likelihood of risk, and address higher risk failure modes for a system (i.e. bicycle), component (i.e. bicycle’s tire), or process (i.e. bicycle assembly) ~\cite{Carlson2012-yh}. FMEA is a multi-step framework, through which steps are iteratively performed by FMEA and system experts over the development life cycle ~\cite{Carlson2012-yh} (see also Fig 1): 

\begin{enumerate}
    \item List out the \textit{functions} of a component/system OR steps of a process (e.g., everything the system/process needs to perform). 
    \item Identify potential\textit{ failure modes}, or mechanisms by which each function or step can go wrong.
    \item Identify the \textit{effect}, or impact of a failure, and score its \textit{severity} on a scale of 1 – 10 (least to most severe).
    \item Identify the \textit{cause}, or why the failure mode occurs, and score its \textit{likelihood of occurrence} on a scale of 1 – 10 (least to most likely). 
    \item Identify \textit{controls}, or how a failure mode could be detected, and score \textit{likelihood of detection} on a scale of 1 – 10 (least to most likely).
    \item Calculate \textit{Risk Priority Number} (RPN) by multiplying the three scores;  higher RPN indicates higher risk level. 
    \item Develop \textit{recommended actions} for each failure mode and prioritize based on RPN. 
\end{enumerate}

\begin{figure}[ht]
  \centering
  \includegraphics[width=\linewidth]{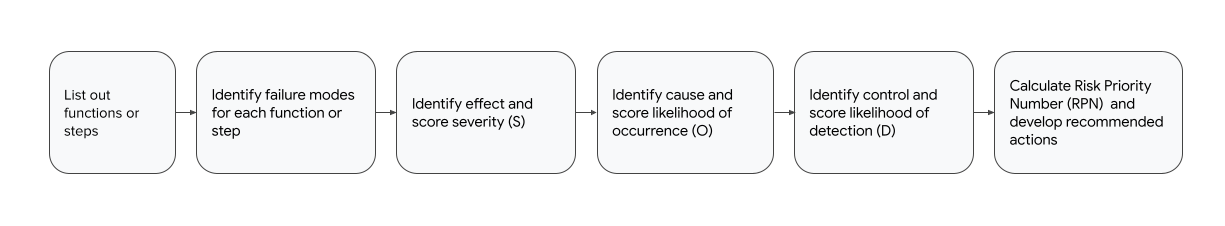}
  \caption{Steps for conducting an FMEA \cite{Carlson2012-yh}}
  \Description{The figure illustrates the most commonly performed steps for FMEA} 
  \label{fig:fmea}
\end{figure}

\subsubsection{System Theoretic Process Analysis (STPA)}
The hazard analysis method, STPA, is a relatively new technique taking a system theoretic perspective towards safety ~\cite{leveson2016engineering}. It maps elements of a system, their interactions, and examines potential hazards (i.e. sources of harm). While analytic reduction requires a user of the tool to imagine interactions between components, modeling at the system level is meant to capture \emph{emergent} phenomena that are well-described only by component interactions rather than individual component behavior. STPA has been employed in NASA’s space program ~\cite{Ishimatsu2014-ni}, the nuclear power industry ~\cite{Shin2021-tg}, and the aviation industry ~\cite{ishimatsu2010modeling}.

In contrast to FMEA, the STPA process does not focus on reliability, failures, or risk likelihood.  Instead, STPA models the sociotechnical system, focusing on the structure between components as well as control and feedback loops. Broadly, STPA encompasses the following steps, which are meant to be iterative (across the model of a system) and cyclic (across a system’s lifecycle) (see Fig 2).
\begin{enumerate}
    \item Define the \textit{purpose of the analysis} by identifying losses via outlining stakeholders, and their values. System specific hazards and controls are then highlighted based on the specified loss.  
    \item Model the \textit{control structure} of the full sociotechnical system using control feedback loops which consists of a controller which sends \textit{control actions} to a system that is being controlled while receiving \textit{feedback} from the same system. 
    \item Identify \textit{unsafe control actions} (UCA) by going through each control action and thinking about unsafe modes of (no) action, incorrect action and untimely action. 
    \item Identify potential \textit{loss scenarios} by outlining potential casual scenarios for each UCA.
\end{enumerate}
These steps can be applied to positive effect at any stage in development, and be used to develop requirements that need to be enforced to ensure a safe sociotechnical system, such as new design decisions, requirements, procedures, operator training, test cases, or even periodic audits.

\begin{figure}[ht]
  \centering
  \includegraphics[width=\linewidth]{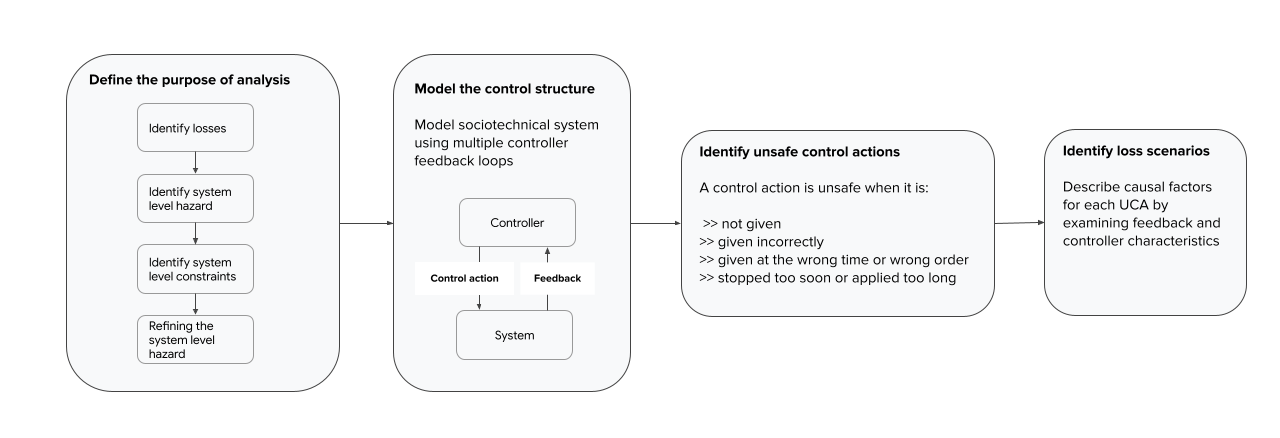}
  \caption{Steps for conducting an STPA\cite{Leveson2018-no}}
  \label{fig:stpa}

\end{figure}

In sum, FMEA and STPA frameworks pose complementary analytical perspectives from safety engineering. Prior work suggests these techniques could strengthen the identification and mitigation of social and ethical risks of \ac{ML} systems ~\cite{Li2022-vt,Dobbe2022-kf,Rismani2021-qy,Raji2020-dw}. Scholars have discussed the overall benefits of FMEA for internal ML auditing ~\cite{Raji2020-dw}, illustrating how it could uncover ML fairness related failures ~\cite{Li2022-vt}, and have used it to propose an analysis of "social failure modes" for ML systems ~\cite{Rismani2021-qy}. Yet, we could not locate any studies investigating ML practitioner’s perspectives towards use of FMEA for social and ethical risk management. Similarly, several works suggest the value of a system theoretic framework for eliminating or mitigating social and ethical risks of ML systems ~\cite{Dobbe2022-kf,Martin2020-dl}. These works illustrate the theoretical application and benefit; however, little work to date explores industry \ac{ML} practitioner’s perspectives towards these techniques and how they could address perceived gaps in current risk management practices ~\cite{Martelaro2022}. 

\section{Methodology}
\label{methodology}

We conducted 30 semi-structured interviews with ML industry practitioners specializing in assessing and mitigating ML ethics risks, from six companies. The research proposal, the interview protocol, and consent forms were reviewed and approved within one of the institutions represented in this study. Here, we describe the participants, recruiting, data collected, analysis, and study limitations.

\subsection{Participants and recruiting}
We used purposive and snowball sampling to recruit participants. Recruitment inclusion criteria specified participants be 18 years old or older, and currently work in an industry position conducting, managing, or researching social and ethical risks of ML system. As our primary research question is to understand industry adoption of reliability engineering tools, we excluded practitioners in academic, governmental or not-for-profit organizations. While we did not establish specific quotas for each professional position, we sought a balance of roles and backgrounds. 

Four of the authors brainstormed an initial list of interview candidates based on knowledge about their existing work profile (via networking and publication or presentation track record at major conferences) and sent emails inviting their participation. Once a candidate accepted an invitation to participate, the interview was scheduled and the interviewer sent the consent form. At the conclusion of each interview session, we invited participants to recommend other candidates. The lead author conducted all interviews, which lasted approximately 60 minutes except for two 90-minute interviews.

In total, 30 practitioners from a diverse range of industry roles and educational backgrounds took part in the study \textbf{(Table\ref{participants}).} Participants held a range of roles including management (e.g., product, technical program, research and executive) (\textit{n}=8), research (\textit{n}=11), analyst/advisory roles (\textit{n}=9), and software engineers (\textit{n}=2). All participants worked in their current role for at least one year, and had experience assessing multiple ML systems for social and ethical risks, including classifiers, recommendation systems, large language models and text to image models. We conducted interviews between June and August 2022. All participants gave informed consent prior to participating in the study; interviews were recorded with permission. Participants were not financially compensated for their participation.

\begin{table}[!ht]
\caption{Participant's roles and reference ID}
\label{participants}
\fontsize{6}{7}\selectfont
\centering
\resizebox{\linewidth}{!}{
\begin{tabular}{p{3cm}p{3.25cm}p{.75cm}p{2cm}}
\hline
\textbf{Job Title} & \textbf{Description} &\textbf{n (\%)} & \textbf{ID}                         \\ \hline
\textbf{Research} (i.e. research scientist, principal researcher)              & Primarily conduct interdisciplinary \par research in responsible ML &11 (37)                             & R3, R4, R5, R12, R18, R19, R21, R25, R22, R28, R29 \\ \hline
\textbf{Analyst/advisory} (i.e. ethics \par reviewer, ethics and policy advisor, sociotechnical analyst, user \par researcher, research associate) & Advise project teams and \par review ML systems according to \par internal review processes &
  9 (30) &
  R6, R7, R8, R13, R14, R16, R17, R24, R26 \\ \hline
\textbf{Management} (i.e. product \par manager, technical program \par manager, research manager, chief \par executive officer) & Manage products, programs, \par companies, and research projects &
  8 (27) &
  R1, R2, R9. R10, R15, R20, R23, R27\\ \hline
\textbf{Engineer} (i.e. research/software \par engineer) & Design and develop ML systems & 2 (6) & R11, R30 \\ \hline
\end{tabular}}
\end{table}

\subsection{Interview design}

The interview protocol consisted of two parts: a) current practices and challenges, and b) first impressions of FMEA and STPA applicability for ML systems. Following confirmation of consent, we asked participants to describe their role and the type of technologies they focus on. We then asked participants how they define, assess, and mitigate social and ethical risk, broadly conceived. Moreover, we asked participants to discuss  challenges they face when assessing and mitigating social and ethical risks in their current role. In the second part of the interview, we introduced the two processes using non-ML examples: FMEA was described with an example of a car tire; STPA was introduced using an example of a new surgical technique. The introduction of each process (including the example) took approximately 5 minutes. We introduced each technique one at a time and then discussed it for 10 minutes each. During this discussion, we asked participants to share their first impressions (pros, cons) while considering their potential use as a social and ethical risks assessment tool for ML systems. We invited them to talk through how they would apply such a process on an ML system they have assessed previously. To avoid order bias, the interviewer alternated between the processes for each interview. All interviews were conducted online using a video conferencing platform. Participants discussed both techniques in all interviews except in two interviews where due to time restraints one of the techniques was not discussed. This occurred once for each of the technique.

\subsection{Data analysis}
We used reflexive thematic analysis \cite{Braun2020-mg, braun2019reflecting} to understand the main themes in the interview data. We used an automatic transcription software for transcribing the interview recordings and then manually cleaned the transcripts. The primary author removed identifying information (e.g., current employer, specific products/projects mentioned) from the transcripts to protect the anonymity of the participants. Four of authors coded the data, first taking familiarization notes to highlight key ideas emerging early in the analysis. We then conducted open coding of the interview data, using the QSR NVivo 12 qualitative analysis software. The lead author coded all of the interviews and three other authors collectively coded 15 interviews. Every interview was coded by two researchers. The authors responsible for coding met iteratively to discuss codes, data interpretations, and progress from codes to thematic discussions. During these discussion session, researchers resolved disagreements, and generated new codes as relevant concepts emerged. In the final session, these authors convened to organize codes thematically and discussed emerging themes. The lead author compiled all the coding documents and synthesized the themes from the group discussions. Next, thematic findings were shared with broader research team for confirmation and collaborative discussion.


\subsection{Author reflexivity}
As with all research, our positionality and lived experiences inform our approach to designing, conducting, and analyzing this research study.
All authors are researchers living in Canada and the United States. Our collective disciplinary backgrounds informing our research perspectives include ML research and engineering, mechanical engineering, robotics, human-robot interaction, sociology/ science and technology studies, cognitive sciences, and cybersecurity. 

\subsection{Study limitations}
Our study examines how ML practitioners engage in social and ethical risk management practices, what challenges they face, and how failure and hazard analysis frameworks could inform and improve their practice. As an exploratory study, further work is needed to deepen understanding and develop ML model or other contextually-specific insights on the applicability of FMEA and STPA. Moreover, the ML practitioners interviewed for the study did not have expertise in safety and reliability engineering, and had limited time and exposure to the techniques. This study reflects first impressions of these frameworks based on their experience. In addition, our participants primarily come from larger, multinational technology organizations (4 of 6 companies represented) and reside in North America. As industry practitioners, there are limitations on what some participants could disclose due to confidentiality commitments. Thus, further work could examine views from a wider range of practitioners, which could provide deeper insights. 

\section{Findings}
\label{findings}

Our study examines how failure and hazard analysis frameworks could inform ML risk management practices. We present our findings in two parts and start by highlighting what practitioners identify as existing social and ethical risk management practices and discuss the challenges they remark. In section two, we build on this understanding of current practices and challenges and discuss ML practitioner’s impressions on how FMEA- and STPA-like processes could strengthen existing practices.

\subsection{Current social and ethical risk management practices}
Participants described increased formalization of risk management practices, yet noted key aspects of their work - including defining and assessing for social and ethical risks - were characterized by an interpretive flexibility through which practitioners navigate peers with multiple and sometimes conflicting understandings of risk management work. While this flexibility accommodates the wide range of ML systems and contexts of deployment  these practitioners are responsibilized to assess, it also fosters friction in multidisciplinary environments. Organizational culture and resource constraints are power dynamics influencing these challenges.
 
\subsubsection{Variable definitions of risk}
\label{definition}
Defining social and ethical risks sets the bounds of which system (mis)behaviors or downstream effects are acceptable or concerning. Rather than anchor to a canonical definition, we find participants employ multiple definitions of social and ethical risks, explicitly noting there is no widely-accepted definition in the ML community. Despite lack of common definitions, there were points of convergence, each underpinned by concerns with adverse, material impacts to people. Foremost, participants described social and ethical risks as \textit{user and societal harms of ML systems}. Here, participants described \textit{“harms"} broadly, without specifying uniform methods for surfacing harms to who or what. While some participants noted a general \textit{“user-centric [harms] framing works well [...] for a product and engineering organization”} (R15), others centered harms to “underrepresented” (R13) and “historically marginalized” (R4) communities. Beyond harms, participants also described how \textit{transgressions to a company's public AI ethics principles} offer a \textit{“jumping off point”} (R15) to identify social and ethical risks. While these commitments provided a clear north star for identifying risks, participants also noted their abstracted nature does not help them know which impacted stakeholders to prioritize or how to grapple with the constitutive role of context-of-use and system affordances (R15) in generating risks. Lastly, participants described social and ethical risks as \textit{human rights violations }that could be surfaced through human right impact assessment. The human rights frame was less common than other definitions, though many participants recognized its value in evaluating systems deployed in cross-cultural domains.

While variable definitions allow for flexibility in scoping social and ethical risks for a wide range of ML systems, they can also foster frustration, misunderstanding, and inefficiency in multidisciplinary environments. Participants all stated clear definitions are necessary for productive conversation (R1, R10), for instance, as (R1), a product manager illustrated: \textit{“sociologists come from the harm perspective, whereas engineers often think of it in the failure perspective.”} Without common language, identification and assessment of risks can be slowed. Yet, multidisciplinary expertise is highly desirable. Some participants in engineering roles reflected they felt ill-equipped to define risks (R11) and expressed need for formalized guidelines. Despite desire for clearer definitions, many participants noted value in definitional flexibility, particularly for assessing novel technologies, in which strict definitions may not accommodate possible harms.

\subsubsection{Multiple methods for assessing social and ethical risks}
\label{assessment}
Whereas definitions of social and ethical risk constitute the boundaries of (un)acceptable ML system behavior, methods shape the situated assumptions, guide the questions asked, and format how social and ethical risks are communicated. Participants described employing various risk assessment methods including qualitative, quantitative, and "reflexive investigatory" approaches, through which the methods and motivation of fellow practitioners are probed for alignment with organization principles and best practices. Participants from four of the six companies indicated they have formal ethical review teams or programs, through which structured risk assessment occurs. Even within these formal structures, however, participants may combine or tailor methods employed based on identified needs.

Consistently, ML practitioners noted they begin social and ethical risk assessment by \textbf{qualitatively mapping potential harms of an ML system} individually or when possible, in teams with interdisciplinary expertise (i.e. product managers, AI ethicists, software engineers, and research scientists working collaboratively). Mapping harms focuses attention on the adverse material impacts to people, including how ML systems can change work practices, socialization patterns, and other dimensions of social life (R18). Mapping harms involves \underline{\textit{surveying existing literature}}, with a focus on known impacts of\textit{ “related technologies”} and social contexts (R25); although, many participants, especially those whose jobs did not include a significant research mandate, expressed time and resource constraints limit their ability to engage literature as deeply as desired. Literature surveys may incorporate non-scholarly perspectives, as one research scientists noted they centered community disability justice perspectives (R5). Mapping harms also involve \underline{\textit{foresight exercises}} to hypothesize potential worst/best case scenarios (R16) through free-form brainstorming and by working through structured questions created internally. Participants also referred to using more formalized assessment process such as human rights and impact assessment processes \cite{Moss2021-qi,Latonero2021-fl} at this stage where the focus is to qualitatively evaluate potential implications. A third, and highly desired approach by participants, when resources permit, is \underline{\textit{participatory methods}}, where community-based stakeholders are engaged to co-identify social and ethical risks. 

Engineers and computer scientists also described \textbf{quantitatively testing ML system properties}. Assessment begins with functional tests, where \textit{“ML components are treated much like a piece of software”} (R11) and subject to routine code reviews and performance tests measuring accuracy, recall, and precision. Functional tests do not explicitly measure social and ethics risks. Assessments for such risks are additive and \textit{“bespoke for every project”} (R11), which may include disaggregated analysis \cite{andrus2021we, barocas2021designing}, counterfactual and causal analysis \cite{garg2019counterfactual}, and adversarial testing \cite{ettinger2017towards,ruiz2022simulated, zeng2021openattack, zhang2020adversarial}. These assessments are conducted pre- and post-launch, and aim to identify allocative, representational, and quality-of-service harms based on identity characteristics. They cannot, as participants note, capture non-computational harms, particularly the diffused and long term impacts of ML systems in the world. Moreover, participants described limitations in post-launch assessment, as there are no rigorous ways of identifying such risks unless reported by users, media outlets, or external auditors.

Lastly, participants in management roles described \textbf{interrogating product and research development processes}. This approach is motivated by participants' recognition that technologies have a world view influenced by the norms, intentions and common practices of researchers and developers. Participants described reviewing product team documentation and methodologies, and making recommendations to improve practices to minimize harms to marginalized communities, such as assessing how a product team ML model evaluation process and identifying whether they are operationalizing any responsible ML metrics \cite{Rakova2021-ov}. For instance, R24, an ethicist, described \textit{“assess[ing] the intentions of teams and … predict[ing] … the[ir] impacts.”} While participants did not detail how they conduct such epistemological assessments, they did reference external literature, including value analysis \cite{Friedman2019-jp}. 

Overall, practitioners noted tensions in risk assessment. First, assessment is most effective when there is commitment to multidisciplinary collaboration between product teams and “subject matter experts” (i.e. non-engineering practitioners, such as ethicists, sociologists, or people with contextual expertise), although such collaboration is \textit{“hard”} given different epistemological background (R19). Second, assessments are \textit{“very product dependent”} in which many participants remarked on the need to better standardize. However, assessment cannot be easily automated as it requires meaningful \textit{“conversations with the product team to understand”} the product, its use case, and where harms may arise (R6).

\subsubsection{Emerging approaches for mitigating social and ethical risks}
\label{mitigation}
Social and ethical risks are often surfaced by ethicists and social scientists who sit outside of research and product teams, and are not subject to product launch incentive structures. As such, mitigating identified risks requires significant work to build cross-functional “partnerships” (R16) and gain buy-in from teams with relevant technical expertise and control to adjustment models and product design. 

Product managers may take charge in mapping a mitigation strategy, however, deciding on an approach also requires collaboration, as preferred strategies vary by disciplinary training. Engineers often gravitated towards algorithmic solutions, such as fine-tuning model parameters, creating new training datasets, and implementing blocklists or filters \cite{Ngo2021-jj} to prevent harmful model inputs or outputs. In contrast, ethicists, social scientists, and designers emphasized UX solutions, policy development, explainability and transparency artifacts, and education. Yet, practitioners all recognized need for multiple interventions, as one computer scientist elaborated:
\begin{quote}
\textit{“...I tend to gravitate towards algorithmic solutions [...but ] want to qualify this is not the only way to solve things [and] there are some things … not mitigatable by algorithmic techniques. In which case, essentially, I defer my expertise to somebody else because maybe the solution in that case, is more on the policy side or participatory design methods outside the scope of what I'm familiar with.” (R3)}
\end{quote}
Prioritizing mitigations is also a challenge, as some recommendations may \textit{“take months, maybe even years to fully fix.”} (R8). There are no clear guidelines on what mitigations need to happen and which ones can be put on hold; though some noted movement towards formalizing mitigation frameworks (R16). As such, resource availability and product team’s priorities dictate which mitigations will be pursued.

\subsection{Challenges in current social and risk assessment approaches}
Our findings show the emergent and chaotic nature of existing practices. We identify four key challenges that ML practitioners face when developing and implementing these practices. 

\subsubsection{Organizational structure: Incentive conflicts, fractured adoption, and unclear responsibilization}
Participants emphasized organizational incentive structures complicate identifying and mitigating social and ethical risks. Research teams are typically incentivized to publish academic papers, and product teams are incentivized to launch new products and features. As such, incorporating social and ethical risk practices can be perceived as slowing down publications and launches. In this way, organizational culture and priorities are a constitutive factor in assessing and mitigating risks—especially those that may take extensive time and resources to fix, as one research scientist, explains:
\begin{quote}
\textit{“When we think about [social and ethical risks of] machine learning, it's not just the algorithms and the code and the data. It also bleeds into the organizational structure, which means that for you to push for a mitigation strategy, internally, or even an assessment strategy, internally, it needs to be recognized as something of value. Because a product team, especially, is going to say, ‘Well, this is nice, but it doesn't align with how I'm being measured in terms of my team's performance or my individual performance.’” (R3)}
\end{quote}

Similarly, R17, an AI ethics analyst, notes \textit{"I think inherently, what we do is not aligned with a corporation … it's not revenue generating work. It's work that can inhibit the bottom line and a product launch. … it can be hard to get product teams to mitigate … because they just want to launch the product.”} The often partial—or non-systemic—ways organizations adopt social and ethical risk management is thus a challenge. Organizations may have formal ethical review teams; yet, if these are “opt-in” rather than mandatory, adoption is fractured or ad hoc.

As the development and use of ML system components are typically diffuse, and often lack centralized control, it may not be clear which job functions are responsible for a given component. For instance, R1, a product manager, reflects on how identifying who is responsible for assessing components is not always straightforward: \textit{“Who is responsible for [assessing the representativeness of datasets used for adversarial testing of downstream products and features]? [...] is the research team responsible or is it the product managers [for where that dataset is used]?”} As well, the team assessing a system component may not \textit{“have decision-making power”} (R17) to act on findings, creating barriers to mitigating identified risks. This raises questions about who is responsible for implementing these processes into research and product development, and who is responsible for establishing the minimum standards researchers and product teams should meet. One program manager emphasized some teams \textit{“are already trying to practice on their own without guidance and really struggle with "How do I make decisions? What's fair enough?”} and expressed concern that, \textit{“If I'm wrong, am I liable?”} (R2). Thus, in recognizing incentive conflicts, fractured adoption, and unclear responsibilization, practitioners often expressed desire for clear guidance on assessing ML systems.

\subsubsection{Resource constraints: time, capacity, data}
Practitioners motivated to address social and ethical risks encounter time, capacity, and resource constraints. For instance, testing and remediating ML systems requires creation of datasets (e.g., for re-training, fairness benchmark testing, etc.). Responsible dataset development is time consuming ~\cite{Paullada2021-eu} and many practitioners are pressured to complete this work in time-constrained environments. R24, who works as a research manager, elaborates the challenges of creating datasets for adversarial product testing under time constraints, which may not leave room for critical self-reflection: 

\begin{quote}
\textit{“It is always very difficult because it is an expedited research project. We have to do it in a month … and every time you have to create a dataset [and] you literally have to sit down and think about all the keywords that you were going to put in a model that could go wrong and ultimately, you're creating a data set, but you’re creating a dataset at a point in time and we know that people have all these cognitive biases.”}
\end{quote}
Similarly, R12, who works as research scientist, describes how dataset creation involves forecasting and anticipating context:
\begin{quote}
\textit{
“A lot of the challenges are obtaining datasets that we think are representative of the downstream deployment context that we think a model is going to end up in. [...] if you are actually interested in mitigating risks that might fall predominantly on certain social groups, you actually need to collect data on identifying those social groups and that is often a difficult thing to do. Alternatively, for some mitigation strategies, you often need to collect very rich data in order for those mitigation strategies to work and many datasets we have were not collected with the intention of building predictors that would actually transfer well in situations where the composition of social groups changes or something like that. And had you known beforehand … we might have been able to guide the data collection a little bit better.”}
\end{quote}
Whether a dataset can be characterized as a “good” dataset is context dependent, and requires foresight, input early in the data collection, and necessary expertise to be able to identify what needs to be collected. Creating appropriate dataset is only an example of how resource limitations that practitioners face when evaluating and mitigating social and ethical risks. Similar challenges were noted for the practices highlighted in Sections \ref{assessment} and \ref{mitigation}.

\subsubsection{Who’s in the room? The need for more diverse perspectives and forms of expertise}

Participants shared concerns how the people in the room define what ethical and social risks are, which is often inadequate and limiting. Speaking directly to this, R25, a research scientist, emphasized:\textit{ “One of the biggest challenges when we're trying to assess and mitigate risks is: Who is it in a room?”} Similarly R17, an AI ethicist reflected: \textit{“[...] we have inherent blind spots on the team. When we think about our lived experience, we do have diverse backgrounds, but ultimately we can't speak to a global perspective as [much] we might like. And when we're imposing an American lens … it's not as robust an analysis as it could be.”} Although practitioners noted they\textit{ “were actively trying to address” }(R25) this problem, actually doing so can be perceived as having high time and effort costs, which can conflict with organizational incentives. As R15, a product manager, explains:
\begin{quote}
\textit{“Whose expertise do you need to do a good job at assessing ethical and social risks for a project? Generally, those people aren't going to be the people who are on your team and therefore it's costly in terms of time and potentially these sources to find those people and to get them in the room and to get them up to speed, educate them on what you're trying to do and then actually get them to help you brainstorm the risks.”  
}
\end{quote}
However, even identifying what is a relevant context-of-use can be challenging, especially as ML systems are deployed globally and  across numerous contexts at once. 

\subsubsection{Uncertainty and knowledge gaps for assessing ML systems}
Participants described different “knowledge gaps” that pose challenges to social and ethical risk assessment. Foremost, participants noted they anticipate and mitigate risks for emerging and novel technologies, which have not yet been deployed in the world. This increases practitioners' uncertainty in forecasting relevant risks. As R8, an AI ethicist, describes, \textit{“everything we see is at the cutting edge of technology. That’s exciting and it also means there is no road map.”} Similarly, R4, a research scientist, emphasized the challenge of \textit{“unknown unknowns,”} reflecting \textit{“How can I find out different kinds of social and ethical risks that might exist that I just am currently unaware of?”} Yet, practitioners highlighted knowledge gaps around established ML technologies, especially in connecting identified risks to relevant system components. R25, a research scientist, elaborates:\textit{ “one of the biggest open problems is, how do you operationalize these complex social constructs in a way that enables us to actually scale evaluation?”}

A second type of knowledge gap reflects the uncertainty of assessing ML systems that are often opaque and complex, through which it is not easy to pinpoint why a ML model is making a particular decision, nor is it easy to operationalize measurement of social and ethical impacts. Participants noted the complexity of ML systems pose challenges to identifying what even constitutes a system. R12, a research scientist, notes \textit{“One thing from my experience that I found is that it's very difficult at least in ML systems to get people to write down a model of the entire system.”} This knowledge gap is especially salient when conducting assessments that require scoring severity or likelihood, which often require forecasting and hypothesizing. R22, a research scientist, describes: \textit{“Uncertainty is a really big [challenge]. Sometimes [ethical and social] concerns are serious. But, it's not easy to parameterize things in the way a risk suggests. So you don't know if it's 10\% likely, 90\% likely. You just have deep uncertainty about whether a future risk will transpire. Particularly, when we talk about [compounding] effects and complex systems.”} Similarly, R7, a research associate, describes\textit{ “we don't have that faster and harder feedback loop on our actions”} and it is unclear what the \textit{“marginal impact”} of an ML based project is at the societal level. Grappling with uncertainty of existing and emerging ML systems is top of mind for practitioners responsible for assessing and mitigating potential social and ethical risks.

\subsection{First impressions on safety engineering frameworks}
Participant first impressions of FMEA and STPA underscored how  safety engineering could bring greater structure to social and ethical risk management practices. However, practitioners emphasized understanding the context of use and implementation remains a critical aspect of assessment, and raised concerns about employing safety engineering when context is not yet known. Other identified limitations include conflicts with organizational norms and capacity. 

\subsubsection{FMEA- and STPA-like processes provide sound structure}
As participants clearly articulated the need for formalized social and ethical risk management guidance, they strongly agreed FMEA- and STPA-like processes provide valuable systematic and structured guidelines for examining potential social and ethical risks for ML systems. When asked about their initial impression of FMEA and its applicability to ML systems, one research scientist describes, \textit{“there are definitely cases where explicitly defining failure modes, trying to have a sense of what the potential causes are and how to mitigate them is a really useful framework for machine learning systems and one that has not been terribly well formalized until now.”} (R12). Many appreciated the ordered steps of FMEA and stated thinking about functions/steps, failure modes, effect, cause and detection can effectively guide existing foresight exercises that evaluators perform to assess social and ethical risks. A few noted similarities between the FMEA process and current foresight exercises they employ. 

Similarly, when discussing the STPA framework, participants noted the system theoretic approach is valuable because it frames analysis of an ML system in relation to both co-existing ML systems and societal power structures. An analyst expresses \textit{“I feel it provides a really sound structure to a process that we need in machine learning systems, especially from the ethical analysis perspective”} (R17). Practitioners emphasized the structure of STPA can support examining the complex interconnections between systems. A research scientist elaborates this point: \textit{“I do think a systems theory approach is very useful. It helps [an evaluator] understand relations between new pathways of harm and allows [them] to  think about the multiple points of intervention”} (R21). Overall, participants noted STPA and FMEA provide complementary and different analytical perspectives for examining failures and hazards. 

\subsubsection{Understanding context is critical for social and ethical risk management: FMEA and STPA have limitations}
In reflecting on the efficacy of these safety engineering processes, participants noticed FMEA does not provide a framework for systematically thinking about the human-ML system interactions when mapping ML system’s functions, steps and failure modes. To address this shortcoming, many suggested an FMEA analysis needs to be accompanied by a deep understanding of social issues relevant to a given ML system. As a research scientist explains, FMEA is \textit{“very agnostic to the socio-technical context at first glance and it will be important to outline the use case when thinking about each component or process”} (R21). Similarly, a product manager remarked if the ethics analysts understand the\textit{“context of deployment”} (i.e. where a model will be deployed in a product or feature), they can think of failures that are not just\textit{“component specific”} (R1). Once failure modes are mapped, participants stated that thinking through effect, cause and control can help practitioners gain \textit{“foresight”} (R19) on social and ethical risks.  

STPA, on the other hand, was perceived as a process that considers an ML system in relation to stakeholders and other automated systems that interact with it. A research scientist noted that STPA would be useful to analyze how a ML system \textit{“fits into a larger decision-making process”} (R12). Many participants appreciated that STPA starts from understanding stakeholders, values and losses and provides a framework for mapping relations between humans and the ML systems. A program manager expressed that STPA \textit{“magnifies the fact that when you think of harms you have to have both the technical and then the social and ethical lenses”} and that it is valuable that STPA provides a framework to represent \textit{“all of that”} (R2). Although STPA facilitates incorporates understanding of context via system theoretic perspectives, participants expressed that mapping multidimensional interactions between an ML system and various stakeholders using bidirectional control feedback loops will at times lead to inaccurate depiction of a sociotechnical system. As R24, an ethicist elaborates \textit{“once you start system theoretic analysis, you're going to abstract away and choosing proxies for social phenomena and they're going to be insufficient” }and noted that STPA-like processes should \textit{“go hand-in-hand with expertise”} of understanding \textit{“the limitations of systems processes analysis.”}

Noting the importance of understanding context of use when understanding social and ethical risks, participants questioned and expressed concern about how FMEA- and STPA-like processes would work when a ML system has no defined application or many different ones, such as generative text to image or large language models. 

\subsubsection{Implementation of FMEA- and STPA-like processes require internal capacity building and organizational shifts.}

Many of our participants’ first reactions to  the FMEA and STPA processes was that the current industry culture and lack of internal capacity within technology companies will be hindrances to their adoption. Without a clear demonstration of their usefulness, industry adoption of similar processes will be slow. This sentiment is in line with the current challenges expressed about existing practices. R15, a product manager, states:
\begin{quote}
\textit{My biggest reaction is that [these processes] are so far from where our engineering culture is at. It feels like you would need to hire an entirely new type of person into these companies and over time completely change roles [...]. If [we] want engineering teams to do this themselves or be directly involved in the risk assessment [we] need to dramatically change the incentive structure."
}\end{quote}

The need for organizational culture shift was expressed with respect to both FMEA and STPA. However, a program manager explains, STPA \textit{“will require greater organization across teams and subject matter experts”} (R2) considering its focus on interaction between different systems. Recognizing the challenge of creating these organizational shifts, participants were interested in exploring STPA only if they could see some concrete evidence of how it worked and what it delivered. STPA, in particular, was seen to require heavier internal capacity building to implement successfully, since many participants have some familiarity with FMEA-like processes already while elements of STPA remain foreign. A research scientist explains:

\begin{quote}
\textit{“People often think very linearly and there's a challenge of trying a systems approach. [Practitioners] want to know X causes Y, causes Z, and they want to mitigate right at one of those points [similar to an FMEA], rather than thinking about all the connections between X, Y and Z and what pathway is causing the most harm... [T]he first step is to get people to realize there are multiple relationships between X, Y and Z.”
}\end{quote}
As building capacity requires time and buy-in from different teams with different incentives, participants emphasized it is \textit{“important for these processes to be simple”} (R9) so a diverse group of people can engage them. Practitioners will need to learn new concepts and it will be important to translate terminology used in FMEA and STPA for ML applications. 

\subsubsection{Observations on step-by-steps application of FMEA and STPA}
When asked to walk through the FMEA and STPA processes as illustrated in Figures \ref{fig:fmea} and \ref{fig:stpa}, participants discussed how they would apply each step on an ML system. Many participants remarked they would benefit from using FMEA and STPA processes \textit{“earlier in development”} (R25) when they are tasked with assessing potential social and ethical risk. 

\paragraph{Observations of the FMEA steps}
\begin{itemize}
    \item \textbf{	Identify function or steps:} Practitioners expressed that there is value in outlining functions or steps for an ML system. They suggested either breaking down functions based on intended uses of an ML-based feature or product (i.e. \textit{“to log food, to inform users and to provide the act of tracking.”} (R10)) OR identifying steps for sub-processes along the ML development pipeline (i.e. \textit{“dataset development, annotation, training, evaluation or deployment”} (R25)). Some remarked that it would be challenging to break down functions for the ML model itself. According to a research scientist\textit{ “key features/functions [of an ML system] are often embedded in some distributed representation in the model. This is especially true for larger models, and it is very hard to assess because the boundaries are not there anymore”} (R3).
    \item \textbf{Identify failure mode:} Participants were able and skilled in articulating failure modes (i.e. \textit{“unfavorable chatbot responses to zip codes that are identified as lower socioeconomic status”} (R17)) but expressed uncertainty about their ability to comprehensively identify all potential failure modes due to the emerging nature of ML technologies. Moreover, they stated ethical and social risk often emerge from complex and non-tangible failures which are hard to identify and often need in-depth analysis of the ML system in context of use.
    \item \textbf{Identify effect:} Participants emphasized the importance of identifying the effect on whom or what, and expressed this should be incorporated into the FMEA process. They raised questions and seeked guidance about the extent to which practitioners should be responsible for protecting the interest of the company deploying/developing an ML system as opposed to interests of directly impacted users or the society at large. 
    \item \textbf{Identify cause and control:} These two steps were seen as valuable, and participants noted ML technology companies have some control over changing design/implementation based on identified causes and controls. Participants noted that causal analysis and developing controls for social and ethical risk are active areas of research and suffer from similar challenges described in 4.1.4 about the uncertainties and knowledge gaps in assessing ML systems. 
\end{itemize}

\paragraph{Observations of the STPA steps}
\begin{itemize}
    \item \textbf{Identify purpose of analysis:}  Practitioners in different roles appreciated the start from stakeholder, values and losses. A participant explains \textit{``I like the idea of starting with the negative outcomes, it's much more user oriented at the beginning, in terms of how it impacts them''} (R1). Social scientists and ethicists noted that in-depth value analysis and normative guidance is required in this step.
    \item \textbf{Create control structure:} Participants identified two scopes of analysis for drawing a control structure: internal company processes OR human-ML product interactions. Some noted that it is difficult to set meaningful boundaries and questioned how one could create a control structures for sociotechnical harms such as ecological harms. A participant noted that \textit{“control structure would be very helpful in terms of limiting rather than constantly overextending where all of the potential problems or risks can come from” }(R10). Guidance is needed for where the system needs to be bounded and future work can use lessons from current STPA literature/practice. \cite{Leveson2018-no} Similar to reasons expressed for an FMEA, participants stated it is challenging to create a control structure for an ML model.
    \item \textbf{Unsafe control actions:}  Participants appreciate \textit{“the quadrant logic”} (R10) (i.e. four conditions outlined by the STPA process) for identifying how control actions could be unsafe. Some participants remarked that unsafe control actions could be mapped to \textit{“design choices”} (R25) and stated that it would be valuable to have this type of analysis earlier on in the development process when creators can critically think about unsafe control actions to inform system requirements and corresponding design choices.
    \item \textbf{Loss scenarios:} Similar to FMEA, participants noted it would be hard to set boundaries for identifying causal scenarios for unsafe control actions. Considering the time and design limitations in our interview, participants did not have the time to provide adequate feedback on this step. Further investigation is necessary to understand how this step could be operationalized for ML systems. 
\end{itemize}
\section{Discussion}
\label{discussion}
Policymakers and critics call for establishing strong accountability practices that responsibilize people and organizations for the risks attendant to ML systems~\cite{kroll2017penn, cooper2022accountability}. Strong accountability enables flexibility and experimentation while providing assurance and potential recourse to affected people~\cite{kroll2018fallacy}. A common question related to responsibilizing individuals or organizations is whether a given harm or problem was adequately \emph{foreseeable}. Although failure is often viewed as inevitable, or even desirable, in machine learning~\cite{perrow1984normal, selbst2020negligence}, safety engineering frameworks, such as FMEA and STPA, provide systematic processes to better anticipate risks ~\cite{leveson2016engineering,Carlson2012-yh}. We discuss challenges and opportunities for future work to improve existing social and ethical risk management practice for ML systems.

\subsection{Failure and hazard analysis frameworks are necessary but not enough}
Our findings illustrate that despite increasing formalization of social and ethical risk management, current practices are disjointed and do not follow a systematic process. Without standard processes, it is difficult for responsible entities to systematically identify risks ~\cite{Raji2020-dw}. Moreover, lack of guidelines for assessment and mitigation of social and ethical risk, results in risk ownership without strong accountability, which creates uncertainty and frustration among practitioners ~\cite{Rakova2021-ov}. Uncertainty about appropriate risk management practices and extant organizational challenges prevent creation of enforcement mechanisms that might foster trust among potentially harmed persons and groups ~\cite{Metcalf2019-eo}. To enable emerging policies and regulations ~\cite{NIST,European_Commission2021-mv,Treasury_Board_of_Canada_Secretariat2019-sk}, there is need to create assessments and mitigations of risk into organizational decision points, through which safety engineering could be supportive ~\cite{Dobbe2022-kf}. Based on initial interactions, ML practitioners recognize the value of safety engineering. Practitioners' walk-through of FMEA and STPA frameworks (Section 4.3.4) provides starting points for future exploration, case studies, and testing such techniques. 

However, these frameworks have limited scope of analysis. STPA is designed to be a system safety engineering framework and FMEA is primarily for reliability assurance ~\cite{Leveson2018-no}. These processes cannot adequately answer normative questions such as “is this a good technology for society?” For example, applying FMEA or STPA on an ML application in gender classification could only marginally make the system safer but it will not be able to address concerns with use of automation in gender classification ~\cite{Keyes2018}. Traditionally, FMEA and STPA are applied when there is sufficient understanding of the deployment context (i.e. geographic location, typical user base, etc.). FMEA- and STPA-like analysis require practitioners to make assumptions about a system, which could be wrong and not hold true in different contexts. The outcome of an FMEA and STPA is not always valid across different contexts of application ~\cite{Carlson2012-yh,Leveson2018-no}. Use of STPA and FMEA like frameworks for general purpose ML systems that are deployed at large scale needs further consideration and investigations. ML practitioners who are tasked with assessing and mitigating social and ethical risks need to be aware of and seek to address the identified scoping challenges for techniques such as FMEA and STPA. 

\subsection{Organizational challenges need to be mitigated}
Safety engineering processes, such as STPA and FMEA, hold potential to address some existing challenges with conducting social and ethical risk management - namely, they can provide a comprehensive framework and structure the analysis. However, implementation of such frameworks could suffer if organizational challenges persist, such as insufficient organizational incentives, homogeneous standpoints and perspectives, and lack of resources. The successful adoption of safety engineering in industries such as medical devices and automotive were accompanied by regulatory and organizational transformations ~\cite{styhre2018unfinished}. With movement towards standardization and regularization of the ML industry, it is important to recognize addressing organizational challenges will persist regardless of chosen framework. However, practitioners and company leaders can learn from the growing body of research on operationalizing AI ethics (e.g., ~\cite{Rakova2021}) and conceptualize emerging roles such as an "ethics owner," individuals responsibilized to manage ethical dilemmas within a technology company ~\cite{Moss2020-le}.

\subsection{Research challenges for the ML research community }
Industry practices for addressing social and ethical risks of machine learning are rapidly emerging. Recognizing that safety frameworks are primarily designed to manage technological failures and hazards, we call on the ML research community to engage, study, critique, and improve the existing social and ethical risk management practices. Specifically, we have identified three research foci.

First, as illustrated in our findings, there is a need for clarifying existing conceptualization of social and ethical risks (i.e. harms to user, AI ethics principle transgression and human rights violation) and we posit that developing taxonomies of harm, failure and hazard - distinct concepts in safety engineering-  for adverse social and ethical impacts of ML systems will provide valuable guidance in social and ethical risk management. Existing critical epistemological perspectives ~\cite{Hanna2020-rn, Bardzell2010-ky,Mohamed2020-uu} on defining harms of ML systems should inform such taxonomies. 

Second, as safety frameworks such as STPA and FMEA can benefit from theoretical framing and analytical processes for examining sociotechnical harms. Building on formalized concepts of risk, hazard, harm and failures for social and ethical implications, this work could provide guidance on STPA and FMEA concepts. When conducting an STPA, guidance is needed around what stakeholders, values and losses need to be considered for a given system. Similarly, an FMEA process needs guidance on how to think about the effect and severity of a potential failure mode. Existing participatory frameworks for AI governance ~\cite{Lee2019webuild} and design approaches ~\cite{Friedman2019-jp,Zytko2022,Harrington2022}could inform this research direction.

Lastly, many empirical studies of responsible ML practices have focused on fairness related methods ~\cite{Madaio2020}, transparency artifacts ~\cite{Mitchell2019-ms}, and general AI ethics operationalization issues ~\cite{Rakova2021-ov}. There is a lack of empirical studies for how practitioners are using, adapting and developing social and ethical risk management techniques. More empirical studies are required to validate the applicability, usability, and capability for identifying and managing risks of emerging frameworks across different ML applications and organizational cultures and application cases.

\section{Conclusion}
\label{conclusion}
Challenges with organizational structure, resources constraints, representing diverse perspectives, and uncertainty of assessing ML systems present fertile ground for innovating social and ethical risk management tools. Quantitative, qualitative and reflexive investigative processes are emerging for defining, assessing and mitigating social and ethical risks.  We study existing practices and posit tools from safety engineering could provide value for creating more appropriate frameworks. Our preliminary discussions with ML practitioners about safety engineering frameworks, such as STPA and FMEA, showed these approaches could be adapted to provide the necessary guidance for systematically conducting failure and hazard analysis for social and ethical risks of ML systems. In this work, we discussed the strength and limitations of these two processes and highlighted need for further research. 
\typeout{}
\bibliographystyle{ACM-Reference-Format}
\bibliography{sample-base}


\begin{thebibliography}{130}


\ifx \showCODEN    \undefined \def \showCODEN     #1{\unskip}     \fi
\ifx \showDOI      \undefined \def \showDOI       #1{#1}\fi
\ifx \showISBNx    \undefined \def \showISBNx     #1{\unskip}     \fi
\ifx \showISBNxiii \undefined \def \showISBNxiii  #1{\unskip}     \fi
\ifx \showISSN     \undefined \def \showISSN      #1{\unskip}     \fi
\ifx \showLCCN     \undefined \def \showLCCN      #1{\unskip}     \fi
\ifx \shownote     \undefined \def \shownote      #1{#1}          \fi
\ifx \showarticletitle \undefined \def \showarticletitle #1{#1}   \fi
\ifx \showURL      \undefined \def \showURL       {\relax}        \fi
\providecommand\bibfield[2]{#2}
\providecommand\bibinfo[2]{#2}
\providecommand\natexlab[1]{#1}
\providecommand\showeprint[2][]{arXiv:#2}

\bibitem[NIS(2021)]%
        {NIST}
 \bibinfo{year}{2021}\natexlab{}.
\newblock \bibinfo{title}{{AI} Risk Management Framework | {NIST}}.
\newblock
  \bibinfo{howpublished}{\url{https://www.nist.gov/itl/ai-risk-management-framework}}.
\newblock
\newblock
\shownote{Accessed: 2022-9-10}.


\bibitem[iee(2022)]%
        {ieee}
 \bibinfo{year}{2022}\natexlab{}.
\newblock \bibinfo{title}{{AIS} Standards}.
\newblock
  \bibinfo{howpublished}{\url{https://standards.ieee.org/initiatives/artificial-intelligence-systems/standards/}}.
\newblock
\newblock
\shownote{Accessed: 2022-9-10}.


\bibitem[ISO(2022)]%
        {ISO}
 \bibinfo{year}{2022}\natexlab{}.
\newblock \bibinfo{title}{{ISO} - {ISO/IEC} {JTC} {1/SC} 42 - Artificial
  intelligence}.
\newblock
  \bibinfo{howpublished}{\url{https://www.iso.org/committee/6794475/x/catalogue/}}.
\newblock
\newblock
\shownote{Accessed: 2022-9-10}.


\bibitem[Abid et~al\mbox{.}(2021)]%
        {Abid2021-ls}
\bibfield{author}{\bibinfo{person}{Abubakar Abid}, \bibinfo{person}{Maheen
  Farooqi}, {and} \bibinfo{person}{James Zou}.}
  \bibinfo{year}{2021}\natexlab{}.
\newblock \showarticletitle{Persistent {Anti-Muslim} Bias in Large Language
  Models}.
\newblock  (\bibinfo{date}{Jan.} \bibinfo{year}{2021}).
\newblock
\showeprint[arxiv]{2101.05783}~[cs.CL]


\bibitem[Andrus et~al\mbox{.}(2021)]%
        {andrus2021we}
\bibfield{author}{\bibinfo{person}{McKane Andrus}, \bibinfo{person}{Elena
  Spitzer}, \bibinfo{person}{Jeffrey Brown}, {and} \bibinfo{person}{Alice
  Xiang}.} \bibinfo{year}{2021}\natexlab{}.
\newblock \showarticletitle{What We Can't Measure, We Can't Understand:
  Challenges to Demographic Data Procurement in the Pursuit of Fairness}. In
  \bibinfo{booktitle}{\emph{Proceedings of the 2021 ACM Conference on Fairness,
  Accountability, and Transparency}} (Virtual Event, Canada)
  \emph{(\bibinfo{series}{FAccT '21})}. \bibinfo{publisher}{Association for
  Computing Machinery}, \bibinfo{address}{New York, NY, USA},
  \bibinfo{pages}{249–260}.
\newblock
\showISBNx{9781450383097}
\urldef\tempurl%
\url{https://doi.org/10.1145/3442188.3445888}
\showDOI{\tempurl}


\bibitem[Angwin and Parris(2016)]%
        {Angwin2016-td}
\bibfield{author}{\bibinfo{person}{Julia Angwin} {and} \bibinfo{person}{Terry
  Parris, Jr}.} \bibinfo{year}{2016}\natexlab{}.
\newblock \bibinfo{title}{Facebook Lets Advertisers Exclude Users by Race}.
\newblock
  \bibinfo{howpublished}{\url{https://www.propublica.org/article/facebook-lets-advertisers-exclude-users-by-race}}.
\newblock
\newblock
\shownote{Accessed: 2022-9-3}.


\bibitem[Bahr(2014)]%
        {bahr2014system}
\bibfield{author}{\bibinfo{person}{Nicholas~J Bahr}.}
  \bibinfo{year}{2014}\natexlab{}.
\newblock \bibinfo{booktitle}{\emph{System safety engineering and risk
  assessment: a practical approach}}.
\newblock \bibinfo{publisher}{CRC press}.
\newblock


\bibitem[Ballard et~al\mbox{.}(2019)]%
        {Ballard2019}
\bibfield{author}{\bibinfo{person}{Stephanie Ballard},
  \bibinfo{person}{Karen~M. Chappell}, {and} \bibinfo{person}{Kristen
  Kennedy}.} \bibinfo{year}{2019}\natexlab{}.
\newblock \showarticletitle{Judgment Call the Game: Using Value Sensitive
  Design and Design Fiction to Surface Ethical Concerns Related to Technology}.
  In \bibinfo{booktitle}{\emph{Proceedings of the 2019 on Designing Interactive
  Systems Conference}} (San Diego, CA, USA) \emph{(\bibinfo{series}{DIS '19})}.
  \bibinfo{publisher}{Association for Computing Machinery},
  \bibinfo{address}{New York, NY, USA}, \bibinfo{pages}{421–433}.
\newblock
\showISBNx{9781450358507}
\urldef\tempurl%
\url{https://doi.org/10.1145/3322276.3323697}
\showDOI{\tempurl}


\bibitem[Bardzell(2010)]%
        {Bardzell2010-ky}
\bibfield{author}{\bibinfo{person}{Shaowen Bardzell}.}
  \bibinfo{year}{2010}\natexlab{}.
\newblock \showarticletitle{Feminist {HCI}: taking stock and outlining an
  agenda for design}. In \bibinfo{booktitle}{\emph{Proceedings of the {SIGCHI}
  Conference on Human Factors in Computing Systems}} (Atlanta, Georgia, USA)
  \emph{(\bibinfo{series}{CHI '10})}. \bibinfo{publisher}{Association for
  Computing Machinery}, \bibinfo{address}{New York, NY, USA},
  \bibinfo{pages}{1301--1310}.
\newblock


\bibitem[Barocas et~al\mbox{.}(2021)]%
        {barocas2021designing}
\bibfield{author}{\bibinfo{person}{Solon Barocas}, \bibinfo{person}{Anhong
  Guo}, \bibinfo{person}{Ece Kamar}, \bibinfo{person}{Jacquelyn Krones},
  \bibinfo{person}{Meredith~Ringel Morris}, \bibinfo{person}{Jennifer~Wortman
  Vaughan}, \bibinfo{person}{Duncan Wadsworth}, {and} \bibinfo{person}{Hanna
  Wallach}.} \bibinfo{year}{2021}\natexlab{}.
\newblock \bibinfo{title}{Designing Disaggregated Evaluations of AI Systems:
  Choices, Considerations, and Tradeoffs}.
\newblock
\newblock
\urldef\tempurl%
\url{https://doi.org/10.48550/ARXIV.2103.06076}
\showDOI{\tempurl}


\bibitem[Barocas et~al\mbox{.}(2013)]%
        {Barocas2013-fe}
\bibfield{author}{\bibinfo{person}{Solon Barocas}, \bibinfo{person}{Sophie
  Hood}, {and} \bibinfo{person}{Malte Ziewitz}.}
  \bibinfo{year}{2013}\natexlab{}.
\newblock \showarticletitle{Governing algorithms: A provocation piece}.
\newblock \bibinfo{journal}{\emph{SSRN Electron. J.}} (\bibinfo{year}{2013}).
\newblock


\bibitem[Benjamin(2019)]%
        {Benjamin2019-qt}
\bibfield{author}{\bibinfo{person}{Ruha Benjamin}.}
  \bibinfo{year}{2019}\natexlab{}.
\newblock \bibinfo{booktitle}{\emph{Race After Technology: Abolitionist Tools
  for the New Jim Code} (\bibinfo{edition}{1} ed.)}.
\newblock \bibinfo{publisher}{Polity}.
\newblock


\bibitem[Birhane(2021)]%
        {Birhane2021-kh}
\bibfield{author}{\bibinfo{person}{Abeba Birhane}.}
  \bibinfo{year}{2021}\natexlab{}.
\newblock \showarticletitle{Algorithmic injustice: a relational ethics
  approach}.
\newblock \bibinfo{journal}{\emph{Patterns (N Y)}} \bibinfo{volume}{2},
  \bibinfo{number}{2} (\bibinfo{date}{Feb.} \bibinfo{year}{2021}),
  \bibinfo{pages}{100205}.
\newblock


\bibitem[Blodgett et~al\mbox{.}(2022)]%
        {Blodgett2022-rc}
\bibfield{author}{\bibinfo{person}{Su~Lin Blodgett}, \bibinfo{person}{Q~Vera
  Liao}, \bibinfo{person}{Alexandra Olteanu}, \bibinfo{person}{Rada Mihalcea},
  \bibinfo{person}{Michael Muller}, \bibinfo{person}{Morgan~Klaus Scheuerman},
  \bibinfo{person}{Chenhao Tan}, {and} \bibinfo{person}{Qian Yang}.}
  \bibinfo{year}{2022}\natexlab{}.
\newblock \showarticletitle{Responsible Language Technologies: Foreseeing and
  Mitigating Harms}. In \bibinfo{booktitle}{\emph{Extended Abstracts of the
  2022 {CHI} Conference on Human Factors in Computing Systems}} (New Orleans,
  LA, USA) \emph{(\bibinfo{series}{CHI EA '22}, \bibinfo{number}{Article
  152})}. \bibinfo{publisher}{Association for Computing Machinery},
  \bibinfo{address}{New York, NY, USA}, \bibinfo{pages}{1--3}.
\newblock


\bibitem[Braun and Clarke(2019)]%
        {braun2019reflecting}
\bibfield{author}{\bibinfo{person}{Virginia Braun} {and}
  \bibinfo{person}{Victoria Clarke}.} \bibinfo{year}{2019}\natexlab{}.
\newblock \showarticletitle{Reflecting on reflexive thematic analysis}.
\newblock \bibinfo{journal}{\emph{Qualitative Research in Sport, Exercise and
  Health}} \bibinfo{volume}{11}, \bibinfo{number}{4} (\bibinfo{year}{2019}),
  \bibinfo{pages}{589--597}.
\newblock


\bibitem[Braun and Clarke(2020)]%
        {Braun2020-mg}
\bibfield{author}{\bibinfo{person}{Virginia Braun} {and}
  \bibinfo{person}{Victoria Clarke}.} \bibinfo{year}{2020}\natexlab{}.
\newblock \showarticletitle{One Size Fits All? {{What}} Counts as Quality
  Practice in (Reflexive) Thematic Analysis?}
\newblock \bibinfo{journal}{\emph{Qualitative Research in Psychology}}
  \bibinfo{volume}{18}, \bibinfo{number}{3} (\bibinfo{date}{Aug.}
  \bibinfo{year}{2020}), \bibinfo{pages}{1--25}.
\newblock


\bibitem[Brey(2012)]%
        {Brey2012-pp}
\bibfield{author}{\bibinfo{person}{Philip A~E Brey}.}
  \bibinfo{year}{2012}\natexlab{}.
\newblock \showarticletitle{Anticipatory Ethics for Emerging Technologies}.
\newblock \bibinfo{journal}{\emph{Nanoethics}} \bibinfo{volume}{6},
  \bibinfo{number}{1} (\bibinfo{date}{April} \bibinfo{year}{2012}),
  \bibinfo{pages}{1--13}.
\newblock


\bibitem[Brown et~al\mbox{.}(2021)]%
        {Brown2021-lo}
\bibfield{author}{\bibinfo{person}{Shea Brown}, \bibinfo{person}{Jovana
  Davidovic}, {and} \bibinfo{person}{Ali Hasan}.}
  \bibinfo{year}{2021}\natexlab{}.
\newblock \showarticletitle{The algorithm audit: Scoring the algorithms that
  score us}.
\newblock \bibinfo{journal}{\emph{Big Data \& Society}} \bibinfo{volume}{8},
  \bibinfo{number}{1} (\bibinfo{date}{Jan.} \bibinfo{year}{2021}),
  \bibinfo{pages}{2053951720983865}.
\newblock


\bibitem[Bugalia et~al\mbox{.}(2022)]%
        {Bugalia2022-hr}
\bibfield{author}{\bibinfo{person}{Nikhil Bugalia},
  \bibinfo{person}{Surjyatapa~R Choudhury}, \bibinfo{person}{Yu Maemura}, {and}
  \bibinfo{person}{K~E Seetharam}.} \bibinfo{year}{2022}\natexlab{}.
\newblock \showarticletitle{A systems theoretic process analysis ({STPA})
  approach for analyzing the governance structure of fecal sludge management in
  Japan}.
\newblock \bibinfo{journal}{\emph{Environment and Planning B: Urban Analytics
  and City Science}} (\bibinfo{date}{March} \bibinfo{year}{2022}),
  \bibinfo{pages}{23998083221075639}.
\newblock


\bibitem[Card and Smith(2020)]%
        {card2020consequentialism}
\bibfield{author}{\bibinfo{person}{Dallas Card} {and} \bibinfo{person}{Noah~A
  Smith}.} \bibinfo{year}{2020}\natexlab{}.
\newblock \showarticletitle{On Consequentialism and Fairness}.
\newblock \bibinfo{journal}{\emph{Frontiers in Artificial Intelligence}}
  \bibinfo{volume}{3} (\bibinfo{year}{2020}), \bibinfo{pages}{34}.
\newblock


\bibitem[Carlson(2012)]%
        {Carlson2012-yh}
\bibfield{author}{\bibinfo{person}{Carl Carlson}.}
  \bibinfo{year}{2012}\natexlab{}.
\newblock \bibinfo{booktitle}{\emph{Effective {FMEAs}: achieving safe,
  reliable, and economical products and processes using failure mode and
  effects analysis}}.
\newblock \bibinfo{publisher}{Wiley}, \bibinfo{address}{Hoboken, N.J}.
\newblock


\bibitem[Chen et~al\mbox{.}(2016)]%
        {Chen2016-tb}
\bibfield{author}{\bibinfo{person}{Le Chen}, \bibinfo{person}{Alan Mislove},
  {and} \bibinfo{person}{Christo Wilson}.} \bibinfo{year}{2016}\natexlab{}.
\newblock \showarticletitle{An Empirical Analysis of Algorithmic Pricing on
  Amazon Marketplace}. In \bibinfo{booktitle}{\emph{Proceedings of the 25th
  International Conference on World Wide Web}} (Montr{\'e}al, Qu{\'e}bec,
  Canada) \emph{(\bibinfo{series}{WWW '16})}. \bibinfo{publisher}{International
  World Wide Web Conferences Steering Committee}, \bibinfo{address}{Republic
  and Canton of Geneva, CHE}, \bibinfo{pages}{1339--1349}.
\newblock


\bibitem[Chen et~al\mbox{.}(2021)]%
        {chen2021mandoline}
\bibfield{author}{\bibinfo{person}{Mayee Chen}, \bibinfo{person}{Karan Goel},
  \bibinfo{person}{Nimit~S Sohoni}, \bibinfo{person}{Fait Poms},
  \bibinfo{person}{Kayvon Fatahalian}, {and} \bibinfo{person}{Christopher
  R{\'e}}.} \bibinfo{year}{2021}\natexlab{}.
\newblock \showarticletitle{Mandoline: Model Evaluation under Distribution
  Shift}. In \bibinfo{booktitle}{\emph{International Conference on Machine
  Learning}}. PMLR, \bibinfo{pages}{1617--1629}.
\newblock


\bibitem[Chohlas-Wood et~al\mbox{.}(2021)]%
        {chohlas2021learning}
\bibfield{author}{\bibinfo{person}{Alex Chohlas-Wood}, \bibinfo{person}{Madison
  Coots}, \bibinfo{person}{Emma Brunskill}, {and} \bibinfo{person}{Sharad
  Goel}.} \bibinfo{year}{2021}\natexlab{}.
\newblock \showarticletitle{Learning to be Fair: A Consequentialist Approach to
  Equitable Decision-Making}.
\newblock \bibinfo{journal}{\emph{arXiv preprint arXiv:2109.08792}}
  (\bibinfo{year}{2021}).
\newblock


\bibitem[Cooper et~al\mbox{.}(2022)]%
        {cooper2022accountability}
\bibfield{author}{\bibinfo{person}{A~Feder Cooper}, \bibinfo{person}{Emanuel
  Moss}, \bibinfo{person}{Benjamin Laufer}, {and} \bibinfo{person}{Helen
  Nissenbaum}.} \bibinfo{year}{2022}\natexlab{}.
\newblock \showarticletitle{Accountability in an algorithmic society:
  Relationality, responsibility, and robustness in machine learning}. In
  \bibinfo{booktitle}{\emph{2022 ACM Conference on Fairness, Accountability,
  and Transparency}}. \bibinfo{pages}{864--876}.
\newblock


\bibitem[Corbett-Davies and Goel(2018)]%
        {corbett2018measure}
\bibfield{author}{\bibinfo{person}{Sam Corbett-Davies} {and}
  \bibinfo{person}{Sharad Goel}.} \bibinfo{year}{2018}\natexlab{}.
\newblock \showarticletitle{The measure and mismeasure of fairness: A critical
  review of fair machine learning}.
\newblock \bibinfo{journal}{\emph{arXiv preprint arXiv:1808.00023}}
  (\bibinfo{year}{2018}).
\newblock


\bibitem[Corbett-Davies et~al\mbox{.}(2017)]%
        {corbett2017algorithmic}
\bibfield{author}{\bibinfo{person}{Sam Corbett-Davies}, \bibinfo{person}{Emma
  Pierson}, \bibinfo{person}{Avi Feller}, \bibinfo{person}{Sharad Goel}, {and}
  \bibinfo{person}{Aziz Huq}.} \bibinfo{year}{2017}\natexlab{}.
\newblock \showarticletitle{Algorithmic decision making and the cost of
  fairness}. In \bibinfo{booktitle}{\emph{Proceedings of the 23rd acm sigkdd
  international conference on knowledge discovery and data mining}}.
  \bibinfo{pages}{797--806}.
\newblock


\bibitem[Cramer et~al\mbox{.}(2018)]%
        {Kramer2018}
\bibfield{author}{\bibinfo{person}{Henriette Cramer}, \bibinfo{person}{Jean
  Garcia-Gathright}, \bibinfo{person}{Aaron Springer}, {and}
  \bibinfo{person}{Sravana Reddy}.} \bibinfo{year}{2018}\natexlab{}.
\newblock \showarticletitle{Assessing and Addressing Algorithmic Bias in
  Practice}.
\newblock \bibinfo{journal}{\emph{Interactions}} \bibinfo{volume}{25},
  \bibinfo{number}{6} (\bibinfo{date}{oct} \bibinfo{year}{2018}),
  \bibinfo{pages}{58–63}.
\newblock
\showISSN{1072-5520}
\urldef\tempurl%
\url{https://doi.org/10.1145/3278156}
\showDOI{\tempurl}


\bibitem[D'Ignazio and Klein(2021)]%
        {DIgnazio2021-ls}
\bibfield{author}{\bibinfo{person}{Catherine D'Ignazio} {and}
  \bibinfo{person}{Lauren~F Klein}.} \bibinfo{year}{2021}\natexlab{}.
\newblock \bibinfo{booktitle}{\emph{Data Feminism}}.
\newblock \bibinfo{publisher}{Tantor Audio}.
\newblock


\bibitem[Dobbe(2022)]%
        {Dobbe2022-kf}
\bibfield{author}{\bibinfo{person}{Roel I~J Dobbe}.}
  \bibinfo{year}{2022}\natexlab{}.
\newblock \showarticletitle{System Safety and Artificial Intelligence}.
\newblock  (\bibinfo{date}{Feb.} \bibinfo{year}{2022}).
\newblock
\showeprint[arxiv]{2202.09292}~[eess.SY]


\bibitem[Ehsan et~al\mbox{.}(2021)]%
        {Ehsan2021-st}
\bibfield{author}{\bibinfo{person}{Upol Ehsan}, \bibinfo{person}{Q~Vera Liao},
  \bibinfo{person}{Michael Muller}, \bibinfo{person}{Mark~O Riedl}, {and}
  \bibinfo{person}{Justin~D Weisz}.} \bibinfo{year}{2021}\natexlab{}.
\newblock \showarticletitle{Expanding Explainability: Towards Social
  Transparency in {AI} systems}. In \bibinfo{booktitle}{\emph{Proceedings of
  the 2021 {CHI} Conference on Human Factors in Computing Systems}} (Yokohama,
  Japan) \emph{(\bibinfo{series}{CHI '21}, \bibinfo{number}{Article 82})}.
  \bibinfo{publisher}{Association for Computing Machinery},
  \bibinfo{address}{New York, NY, USA}, \bibinfo{pages}{1--19}.
\newblock


\bibitem[Ehsan et~al\mbox{.}(2022)]%
        {Ehsan2022-gx}
\bibfield{author}{\bibinfo{person}{Upol Ehsan}, \bibinfo{person}{Ranjit Singh},
  \bibinfo{person}{Jacob Metcalf}, {and} \bibinfo{person}{Mark Riedl}.}
  \bibinfo{year}{2022}\natexlab{}.
\newblock \showarticletitle{The Algorithmic Imprint}. In
  \bibinfo{booktitle}{\emph{2022 {ACM} Conference on Fairness, Accountability,
  and Transparency}} (Seoul, Republic of Korea) \emph{(\bibinfo{series}{FAccT
  '22})}. \bibinfo{publisher}{Association for Computing Machinery},
  \bibinfo{address}{New York, NY, USA}, \bibinfo{pages}{1305--1317}.
\newblock


\bibitem[Ericson et~al\mbox{.}(2015)]%
        {ericson2015hazard}
\bibfield{author}{\bibinfo{person}{Clifton~A Ericson} {et~al\mbox{.}}}
  \bibinfo{year}{2015}\natexlab{}.
\newblock \bibinfo{booktitle}{\emph{Hazard analysis techniques for system
  safety}}.
\newblock \bibinfo{publisher}{John Wiley \& Sons}.
\newblock


\bibitem[Ettinger et~al\mbox{.}(2017)]%
        {ettinger2017towards}
\bibfield{author}{\bibinfo{person}{Allyson Ettinger}, \bibinfo{person}{Sudha
  Rao}, \bibinfo{person}{Hal Daum{\'e}~III}, {and} \bibinfo{person}{Emily~M
  Bender}.} \bibinfo{year}{2017}\natexlab{}.
\newblock \showarticletitle{Towards linguistically generalizable NLP systems: A
  workshop and shared task}.
\newblock \bibinfo{journal}{\emph{arXiv preprint arXiv:1711.01505}}
  (\bibinfo{year}{2017}).
\newblock


\bibitem[{European Commission}(2021)]%
        {European_Commission2021-mv}
\bibfield{author}{\bibinfo{person}{{European Commission}}.}
  \bibinfo{year}{2021}\natexlab{}.
\newblock \bibinfo{title}{Proposal for a Regulation laying down harmonised
  rules on artificial intelligence}.
\newblock
\newblock


\bibitem[Feder~Cooper et~al\mbox{.}(2022)]%
        {Feder_Cooper2022-fh}
\bibfield{author}{\bibinfo{person}{A Feder~Cooper}, \bibinfo{person}{Emanuel
  Moss}, \bibinfo{person}{Benjamin Laufer}, {and} \bibinfo{person}{Helen
  Nissenbaum}.} \bibinfo{year}{2022}\natexlab{}.
\newblock \showarticletitle{Accountability in an Algorithmic Society:
  Relationality, Responsibility, and Robustness in Machine Learning}.
\newblock  (\bibinfo{date}{Feb.} \bibinfo{year}{2022}).
\newblock
\showeprint[arxiv]{2202.05338}~[cs.CY]


\bibitem[Floridi and Strait(2020)]%
        {Floridi2020-gu}
\bibfield{author}{\bibinfo{person}{Luciano Floridi} {and}
  \bibinfo{person}{Andrew Strait}.} \bibinfo{year}{2020}\natexlab{}.
\newblock \showarticletitle{Ethical Foresight Analysis: What it is and Why it
  is Needed?}
\newblock \bibinfo{journal}{\emph{Minds Mach.}} \bibinfo{volume}{30},
  \bibinfo{number}{1} (\bibinfo{date}{March} \bibinfo{year}{2020}),
  \bibinfo{pages}{77--97}.
\newblock


\bibitem[Franklin et~al\mbox{.}(2022)]%
        {Franklin2022-qe}
\bibfield{author}{\bibinfo{person}{Jade~S Franklin}, \bibinfo{person}{Karan
  Bhanot}, \bibinfo{person}{Mohamed Ghalwash}, \bibinfo{person}{Kristin~P
  Bennett}, \bibinfo{person}{Jamie McCusker}, {and} \bibinfo{person}{Deborah~L
  McGuinness}.} \bibinfo{year}{2022}\natexlab{}.
\newblock \showarticletitle{An Ontology for Fairness Metrics}. In
  \bibinfo{booktitle}{\emph{Proceedings of the 2022 {AAAI/ACM} Conference on
  {AI}, Ethics, and Society}} (Oxford, United Kingdom)
  \emph{(\bibinfo{series}{AIES '22})}. \bibinfo{publisher}{Association for
  Computing Machinery}, \bibinfo{address}{New York, NY, USA},
  \bibinfo{pages}{265--275}.
\newblock


\bibitem[Friedman and Hendry(2019a)]%
        {Friedman2019-jp}
\bibfield{author}{\bibinfo{person}{Batya Friedman} {and}
  \bibinfo{person}{David~G Hendry}.} \bibinfo{year}{2019}\natexlab{a}.
\newblock \bibinfo{booktitle}{\emph{Value Sensitive Design}}.
\newblock \bibinfo{publisher}{MIT Press}.
\newblock


\bibitem[Friedman and Hendry(2019b)]%
        {Friedman2019-jw}
\bibfield{author}{\bibinfo{person}{Batya Friedman} {and}
  \bibinfo{person}{David~G Hendry}.} \bibinfo{year}{2019}\natexlab{b}.
\newblock \bibinfo{booktitle}{\emph{Value Sensitive Design: Shaping Technology
  with Moral Imagination}}.
\newblock \bibinfo{publisher}{MIT Press}.
\newblock


\bibitem[Friedman et~al\mbox{.}(1994)]%
        {Friedman1994-rm}
\bibfield{author}{\bibinfo{person}{Batya Friedman}, \bibinfo{person}{Nancy
  Leveson}, \bibinfo{person}{Ben Schneiderman}, \bibinfo{person}{Lucy Suchman},
  {and} \bibinfo{person}{Terry Winograd}.} \bibinfo{year}{1994}\natexlab{}.
\newblock \showarticletitle{Beyond Accuracy, Reliability, and Efficiency:
  Criteria for a Good Computer System}.
\newblock \bibinfo{howpublished}{ACM CHI Conference on Human Factors in
  Computing}.
\newblock  (\bibinfo{date}{April} \bibinfo{year}{1994}).
\newblock


\bibitem[Friedman and Nissenbaum(1996)]%
        {Friedman1996-ou}
\bibfield{author}{\bibinfo{person}{Batya Friedman} {and} \bibinfo{person}{Helen
  Nissenbaum}.} \bibinfo{year}{1996}\natexlab{}.
\newblock \showarticletitle{Bias in computer systems}.
\newblock \bibinfo{journal}{\emph{ACM Trans. Inf. Syst. Secur.}}
  \bibinfo{volume}{14}, \bibinfo{number}{3} (\bibinfo{date}{July}
  \bibinfo{year}{1996}), \bibinfo{pages}{330--347}.
\newblock


\bibitem[Garg et~al\mbox{.}(2019)]%
        {garg2019counterfactual}
\bibfield{author}{\bibinfo{person}{Sahaj Garg}, \bibinfo{person}{Vincent
  Perot}, \bibinfo{person}{Nicole Limtiaco}, \bibinfo{person}{Ankur Taly},
  \bibinfo{person}{Ed~H Chi}, {and} \bibinfo{person}{Alex Beutel}.}
  \bibinfo{year}{2019}\natexlab{}.
\newblock \showarticletitle{Counterfactual fairness in text classification
  through robustness}. In \bibinfo{booktitle}{\emph{Proceedings of the 2019
  AAAI/ACM Conference on AI, Ethics, and Society}}. \bibinfo{pages}{219--226}.
\newblock


\bibitem[Gebru et~al\mbox{.}(2021)]%
        {Gebru2021-kx}
\bibfield{author}{\bibinfo{person}{Timnit Gebru}, \bibinfo{person}{Jamie
  Morgenstern}, \bibinfo{person}{Briana Vecchione},
  \bibinfo{person}{Jennifer~Wortman Vaughan}, \bibinfo{person}{Hanna Wallach},
  \bibinfo{person}{Hal~Daum{\'e} Iii}, {and} \bibinfo{person}{Kate Crawford}.}
  \bibinfo{year}{2021}\natexlab{}.
\newblock \showarticletitle{Datasheets for datasets}.
\newblock \bibinfo{journal}{\emph{Commun. ACM}} \bibinfo{volume}{64},
  \bibinfo{number}{12} (\bibinfo{date}{Nov.} \bibinfo{year}{2021}),
  \bibinfo{pages}{86--92}.
\newblock


\bibitem[{Government of Canada}(2022)]%
        {Government_of_Canada2022-bq}
\bibfield{author}{\bibinfo{person}{{Government of Canada}}.}
  \bibinfo{year}{2022}\natexlab{}.
\newblock \bibinfo{title}{Bill {C-27} summary: Digital Charter Implementation
  Act, 2022}.
\newblock
  \bibinfo{howpublished}{\url{https://ised-isde.canada.ca/site/innovation-better-canada/en/canadas-digital-charter/bill-summary-digital-charter-implementation-act-2020}}.
\newblock
\newblock
\shownote{Accessed: 2022-9-10}.


\bibitem[Hanna et~al\mbox{.}(2020)]%
        {Hanna2020-rn}
\bibfield{author}{\bibinfo{person}{Alex Hanna}, \bibinfo{person}{Emily Denton},
  \bibinfo{person}{Andrew Smart}, {and} \bibinfo{person}{Jamila Smith-Loud}.}
  \bibinfo{year}{2020}\natexlab{}.
\newblock \showarticletitle{Towards a critical race methodology in algorithmic
  fairness}. In \bibinfo{booktitle}{\emph{Proceedings of the 2020 Conference on
  Fairness, Accountability, and Transparency}} (Barcelona Spain).
  \bibinfo{publisher}{ACM}, \bibinfo{address}{New York, NY, USA}.
\newblock


\bibitem[Harrington et~al\mbox{.}(2022)]%
        {Harrington2022}
\bibfield{author}{\bibinfo{person}{Christina~N. Harrington},
  \bibinfo{person}{Shamika Klassen}, {and} \bibinfo{person}{Yolanda~A.
  Rankin}.} \bibinfo{year}{2022}\natexlab{}.
\newblock \showarticletitle{“All That You Touch, You Change”: Expanding the
  Canon of Speculative Design Towards Black Futuring}. In
  \bibinfo{booktitle}{\emph{Proceedings of the 2022 CHI Conference on Human
  Factors in Computing Systems}} (New Orleans, LA, USA)
  \emph{(\bibinfo{series}{CHI '22})}. \bibinfo{publisher}{Association for
  Computing Machinery}, \bibinfo{address}{New York, NY, USA}, Article
  \bibinfo{articleno}{450}, \bibinfo{numpages}{10}~pages.
\newblock
\showISBNx{9781450391573}
\urldef\tempurl%
\url{https://doi.org/10.1145/3491102.3502118}
\showDOI{\tempurl}


\bibitem[Holstein et~al\mbox{.}(2019)]%
        {Holstein2019}
\bibfield{author}{\bibinfo{person}{Kenneth Holstein}, \bibinfo{person}{Jennifer
  Wortman~Vaughan}, \bibinfo{person}{Hal Daum\'{e}}, \bibinfo{person}{Miro
  Dudik}, {and} \bibinfo{person}{Hanna Wallach}.}
  \bibinfo{year}{2019}\natexlab{}.
\newblock \showarticletitle{Improving Fairness in Machine Learning Systems:
  What Do Industry Practitioners Need?}. In
  \bibinfo{booktitle}{\emph{Proceedings of the 2019 CHI Conference on Human
  Factors in Computing Systems}} (Glasgow, Scotland Uk)
  \emph{(\bibinfo{series}{CHI '19})}. \bibinfo{publisher}{Association for
  Computing Machinery}, \bibinfo{address}{New York, NY, USA},
  \bibinfo{pages}{1–16}.
\newblock
\showISBNx{9781450359702}
\urldef\tempurl%
\url{https://doi.org/10.1145/3290605.3300830}
\showDOI{\tempurl}


\bibitem[Hope et~al\mbox{.}(2019)]%
        {Hope2019}
\bibfield{author}{\bibinfo{person}{Alexis Hope}, \bibinfo{person}{Catherine
  D'Ignazio}, \bibinfo{person}{Josephine Hoy}, \bibinfo{person}{Rebecca
  Michelson}, \bibinfo{person}{Jennifer Roberts}, \bibinfo{person}{Kate
  Krontiris}, {and} \bibinfo{person}{Ethan Zuckerman}.}
  \bibinfo{year}{2019}\natexlab{}.
\newblock \showarticletitle{Hackathons as Participatory Design: Iterating
  Feminist Utopias}. In \bibinfo{booktitle}{\emph{Proceedings of the 2019 CHI
  Conference on Human Factors in Computing Systems}} (Glasgow, Scotland Uk)
  \emph{(\bibinfo{series}{CHI '19})}. \bibinfo{publisher}{Association for
  Computing Machinery}, \bibinfo{address}{New York, NY, USA},
  \bibinfo{pages}{1–14}.
\newblock
\showISBNx{9781450359702}
\urldef\tempurl%
\url{https://doi.org/10.1145/3290605.3300291}
\showDOI{\tempurl}


\bibitem[Hutchinson et~al\mbox{.}(2022)]%
        {hutchinson2022evaluation}
\bibfield{author}{\bibinfo{person}{Ben Hutchinson}, \bibinfo{person}{Negar
  Rostamzadeh}, \bibinfo{person}{Christina Greer}, \bibinfo{person}{Katherine
  Heller}, {and} \bibinfo{person}{Vinodkumar Prabhakaran}.}
  \bibinfo{year}{2022}\natexlab{}.
\newblock \showarticletitle{Evaluation Gaps in Machine Learning Practice}.
\newblock \bibinfo{journal}{\emph{arXiv preprint arXiv:2205.05256}}
  (\bibinfo{year}{2022}).
\newblock


\bibitem[Ishimatsu et~al\mbox{.}(2010)]%
        {ishimatsu2010modeling}
\bibfield{author}{\bibinfo{person}{Takuto Ishimatsu}, \bibinfo{person}{Nancy~G
  Leveson}, \bibinfo{person}{John Thomas}, \bibinfo{person}{Masa Katahira},
  \bibinfo{person}{Yuko Miyamoto}, {and} \bibinfo{person}{Haruka Nakao}.}
  \bibinfo{year}{2010}\natexlab{}.
\newblock \showarticletitle{Modeling and hazard analysis using STPA}.
\newblock  (\bibinfo{year}{2010}).
\newblock


\bibitem[Ishimatsu et~al\mbox{.}(2014)]%
        {Ishimatsu2014-ni}
\bibfield{author}{\bibinfo{person}{Takuto Ishimatsu}, \bibinfo{person}{Nancy~G
  Leveson}, \bibinfo{person}{John~P Thomas}, \bibinfo{person}{Cody~H Fleming},
  \bibinfo{person}{Masafumi Katahira}, \bibinfo{person}{Yuko Miyamoto},
  \bibinfo{person}{Ryo Ujiie}, \bibinfo{person}{Haruka Nakao}, {and}
  \bibinfo{person}{Nobuyuki Hoshino}.} \bibinfo{year}{2014}\natexlab{}.
\newblock \showarticletitle{Hazard Analysis of Complex Spacecraft Using
  {Systems-Theoretic} Process Analysis}.
\newblock \bibinfo{journal}{\emph{J. Spacecr. Rockets}} \bibinfo{volume}{51},
  \bibinfo{number}{2} (\bibinfo{date}{March} \bibinfo{year}{2014}),
  \bibinfo{pages}{509--522}.
\newblock


\bibitem[Jasanoff(2004)]%
        {Jasanoff2004-fy}
\bibfield{author}{\bibinfo{person}{Sheila Jasanoff}.}
  \bibinfo{year}{2004}\natexlab{}.
\newblock \bibinfo{booktitle}{\emph{States of Knowledge: The {Co-Production} of
  Science and Social Order}}.
\newblock \bibinfo{publisher}{Routledge}.
\newblock


\bibitem[Jenab and Pineau(2015)]%
        {Jenab2015-ts}
\bibfield{author}{\bibinfo{person}{Kouroush Jenab} {and}
  \bibinfo{person}{Joseph Pineau}.} \bibinfo{year}{2015}\natexlab{}.
\newblock \showarticletitle{Failure mode and effect analysis on safety critical
  components of space travel}.
\newblock \bibinfo{journal}{\emph{Manag. Sci. Lett.}} \bibinfo{volume}{5},
  \bibinfo{number}{7} (\bibinfo{year}{2015}), \bibinfo{pages}{669--678}.
\newblock


\bibitem[Jobin et~al\mbox{.}(2019)]%
        {Jobin2019-qf}
\bibfield{author}{\bibinfo{person}{Anna Jobin}, \bibinfo{person}{Marcello
  Ienca}, {and} \bibinfo{person}{Effy Vayena}.}
  \bibinfo{year}{2019}\natexlab{}.
\newblock \showarticletitle{The global landscape of {AI} ethics guidelines}.
\newblock \bibinfo{journal}{\emph{Nature Machine Intelligence}}
  \bibinfo{volume}{1}, \bibinfo{number}{9} (\bibinfo{date}{Sept.}
  \bibinfo{year}{2019}), \bibinfo{pages}{389--399}.
\newblock


\bibitem[Kay et~al\mbox{.}(2015)]%
        {Kay2015-wn}
\bibfield{author}{\bibinfo{person}{Matthew Kay}, \bibinfo{person}{Cynthia
  Matuszek}, {and} \bibinfo{person}{Sean~A Munson}.}
  \bibinfo{year}{2015}\natexlab{}.
\newblock \showarticletitle{Unequal Representation and Gender Stereotypes in
  Image Search Results for Occupations}. In
  \bibinfo{booktitle}{\emph{Proceedings of the 33rd Annual {ACM} Conference on
  Human Factors in Computing Systems}} (Seoul, Republic of Korea)
  \emph{(\bibinfo{series}{CHI '15})}. \bibinfo{publisher}{Association for
  Computing Machinery}, \bibinfo{address}{New York, NY, USA},
  \bibinfo{pages}{3819--3828}.
\newblock


\bibitem[Keyes(2018)]%
        {Keyes2018}
\bibfield{author}{\bibinfo{person}{Os Keyes}.} \bibinfo{year}{2018}\natexlab{}.
\newblock \showarticletitle{The Misgendering Machines: Trans/HCI Implications
  of Automatic Gender Recognition}.
\newblock \bibinfo{journal}{\emph{Proc. ACM Hum.-Comput. Interact.}}
  \bibinfo{volume}{2}, \bibinfo{number}{CSCW}, Article \bibinfo{articleno}{88}
  (\bibinfo{date}{nov} \bibinfo{year}{2018}), \bibinfo{numpages}{22}~pages.
\newblock
\urldef\tempurl%
\url{https://doi.org/10.1145/3274357}
\showDOI{\tempurl}


\bibitem[Klumbyt{\.e} et~al\mbox{.}(2022)]%
        {Klumbyte2022-tf}
\bibfield{author}{\bibinfo{person}{Goda Klumbyt{\.e}}, \bibinfo{person}{Claude
  Draude}, {and} \bibinfo{person}{Alex~S Taylor}.}
  \bibinfo{year}{2022}\natexlab{}.
\newblock \showarticletitle{Critical Tools for Machine Learning: Working with
  Intersectional Critical Concepts in Machine Learning Systems Design}. In
  \bibinfo{booktitle}{\emph{2022 {ACM} Conference on Fairness, Accountability,
  and Transparency}} (Seoul, Republic of Korea) \emph{(\bibinfo{series}{FAccT
  '22})}. \bibinfo{publisher}{Association for Computing Machinery},
  \bibinfo{address}{New York, NY, USA}, \bibinfo{pages}{1528--1541}.
\newblock


\bibitem[Koh et~al\mbox{.}(2020)]%
        {koh2020wilds}
\bibfield{author}{\bibinfo{person}{Pang~Wei Koh}, \bibinfo{person}{Shiori
  Sagawa}, \bibinfo{person}{Henrik Marklund}, \bibinfo{person}{Sang~Michael
  Xie}, \bibinfo{person}{Marvin Zhang}, \bibinfo{person}{Akshay Balsubramani},
  \bibinfo{person}{Weihua Hu}, \bibinfo{person}{Michihiro Yasunaga},
  \bibinfo{person}{Richard~Lanas Phillips}, \bibinfo{person}{Sara Beery},
  \bibinfo{person}{Jure Leskovec}, \bibinfo{person}{Anshul Kundaje},
  \bibinfo{person}{Emma Pierson}, \bibinfo{person}{Sergey Levine},
  \bibinfo{person}{Chelsea Finn}, {and} \bibinfo{person}{Percy Liang}.}
  \bibinfo{year}{2020}\natexlab{}.
\newblock \showarticletitle{WILDS: A Benchmark of in-the-Wild Distribution
  Shifts}.
\newblock \bibinfo{journal}{\emph{CoRR}}  \bibinfo{volume}{abs/2012.07421}
  (\bibinfo{year}{2020}).
\newblock
\urldef\tempurl%
\url{https://arxiv.org/abs/2012.07421}
\showURL{%
\tempurl}


\bibitem[Krafft et~al\mbox{.}(2020)]%
        {Krafft2020}
\bibfield{author}{\bibinfo{person}{P.~M. Krafft}, \bibinfo{person}{Meg Young},
  \bibinfo{person}{Michael Katell}, \bibinfo{person}{Karen Huang}, {and}
  \bibinfo{person}{Ghislain Bugingo}.} \bibinfo{year}{2020}\natexlab{}.
\newblock \showarticletitle{Defining AI in Policy versus Practice}. In
  \bibinfo{booktitle}{\emph{Proceedings of the AAAI/ACM Conference on AI,
  Ethics, and Society}} (New York, NY, USA) \emph{(\bibinfo{series}{AIES
  '20})}. \bibinfo{publisher}{Association for Computing Machinery},
  \bibinfo{address}{New York, NY, USA}, \bibinfo{pages}{72–78}.
\newblock
\showISBNx{9781450371100}
\urldef\tempurl%
\url{https://doi.org/10.1145/3375627.3375835}
\showDOI{\tempurl}


\bibitem[Kroll(2018)]%
        {kroll2018fallacy}
\bibfield{author}{\bibinfo{person}{Joshua~A Kroll}.}
  \bibinfo{year}{2018}\natexlab{}.
\newblock \showarticletitle{The fallacy of inscrutability}.
\newblock \bibinfo{journal}{\emph{Phil. Trans. R. Soc. A}}
  \bibinfo{volume}{376}, \bibinfo{number}{2133} (\bibinfo{year}{2018}),
  \bibinfo{numpages}{14}~pages.
\newblock


\bibitem[Kroll et~al\mbox{.}(2017)]%
        {kroll2017penn}
\bibfield{author}{\bibinfo{person}{Joshua~A. Kroll}, \bibinfo{person}{Joanna
  Huey}, \bibinfo{person}{Solon Barocas}, \bibinfo{person}{Edward~W. Felten},
  \bibinfo{person}{Joel~R. Reidenberg}, \bibinfo{person}{David~G. Robinson},
  {and} \bibinfo{person}{Harlan Yu.}} \bibinfo{year}{2017}\natexlab{}.
\newblock \showarticletitle{Accountable Algorithms}.
\newblock \bibinfo{journal}{\emph{University of Pennsylvania Law Review}}
  \bibinfo{volume}{165} (\bibinfo{year}{2017}), \bibinfo{pages}{633--705}.
\newblock
Issue 3.


\bibitem[Latonero and Agarwal(2021)]%
        {Latonero2021-fl}
\bibfield{author}{\bibinfo{person}{Mark Latonero} {and} \bibinfo{person}{Aaina
  Agarwal}.} \bibinfo{year}{2021}\natexlab{}.
\newblock \bibinfo{booktitle}{\emph{Human Rights Impact Assessments for {AI}:
  Learning from Facebook's Failure in Myanmar}}.
\newblock \bibinfo{type}{{T}echnical {R}eport}. \bibinfo{institution}{Carr
  Center for Human Rights Policy Harvard Kennedy School, Harvard University}.
\newblock


\bibitem[Lee et~al\mbox{.}(2019a)]%
        {Lee2019procedural}
\bibfield{author}{\bibinfo{person}{Min~Kyung Lee}, \bibinfo{person}{Anuraag
  Jain}, \bibinfo{person}{Hea~Jin Cha}, \bibinfo{person}{Shashank Ojha}, {and}
  \bibinfo{person}{Daniel Kusbit}.} \bibinfo{year}{2019}\natexlab{a}.
\newblock \showarticletitle{Procedural Justice in Algorithmic Fairness:
  Leveraging Transparency and Outcome Control for Fair Algorithmic Mediation}.
\newblock \bibinfo{journal}{\emph{Proc. ACM Hum.-Comput. Interact.}}
  \bibinfo{volume}{3}, \bibinfo{number}{CSCW}, Article \bibinfo{articleno}{182}
  (\bibinfo{date}{nov} \bibinfo{year}{2019}), \bibinfo{numpages}{26}~pages.
\newblock
\urldef\tempurl%
\url{https://doi.org/10.1145/3359284}
\showDOI{\tempurl}


\bibitem[Lee et~al\mbox{.}(2019b)]%
        {Lee2019webuild}
\bibfield{author}{\bibinfo{person}{Min~Kyung Lee}, \bibinfo{person}{Daniel
  Kusbit}, \bibinfo{person}{Anson Kahng}, \bibinfo{person}{Ji~Tae Kim},
  \bibinfo{person}{Xinran Yuan}, \bibinfo{person}{Allissa Chan},
  \bibinfo{person}{Daniel See}, \bibinfo{person}{Ritesh Noothigattu},
  \bibinfo{person}{Siheon Lee}, \bibinfo{person}{Alexandros Psomas}, {and}
  \bibinfo{person}{Ariel~D. Procaccia}.} \bibinfo{year}{2019}\natexlab{b}.
\newblock \showarticletitle{WeBuildAI: Participatory Framework for Algorithmic
  Governance}.
\newblock \bibinfo{journal}{\emph{Proc. ACM Hum.-Comput. Interact.}}
  \bibinfo{volume}{3}, \bibinfo{number}{CSCW}, Article \bibinfo{articleno}{181}
  (\bibinfo{date}{nov} \bibinfo{year}{2019}), \bibinfo{numpages}{35}~pages.
\newblock
\urldef\tempurl%
\url{https://doi.org/10.1145/3359283}
\showDOI{\tempurl}


\bibitem[Lee and Singh(2021)]%
        {Lee2021-oo}
\bibfield{author}{\bibinfo{person}{Michelle Seng~Ah Lee} {and}
  \bibinfo{person}{Jat Singh}.} \bibinfo{year}{2021}\natexlab{}.
\newblock \showarticletitle{The Landscape and Gaps in Open Source Fairness
  Toolkits}. In \bibinfo{booktitle}{\emph{Proceedings of the 2021 {CHI}
  Conference on Human Factors in Computing Systems}} (Yokohama, Japan)
  \emph{(\bibinfo{series}{CHI '21}, \bibinfo{number}{Article 699})}.
  \bibinfo{publisher}{Association for Computing Machinery},
  \bibinfo{address}{New York, NY, USA}, \bibinfo{pages}{1--13}.
\newblock


\bibitem[Leveson and Thomas(2018)]%
        {Leveson2018-no}
\bibfield{author}{\bibinfo{person}{Nancy Leveson} {and} \bibinfo{person}{John
  Thomas}.} \bibinfo{year}{2018}\natexlab{}.
\newblock \bibinfo{title}{{STPA\_Handbook}}.
\newblock
\newblock


\bibitem[Leveson(2016)]%
        {leveson2016engineering}
\bibfield{author}{\bibinfo{person}{Nancy~G Leveson}.}
  \bibinfo{year}{2016}\natexlab{}.
\newblock \bibinfo{booktitle}{\emph{Engineering a safer world: Systems thinking
  applied to safety}}.
\newblock \bibinfo{publisher}{The MIT Press}.
\newblock


\bibitem[Li and Chignell(2022)]%
        {Li2022-vt}
\bibfield{author}{\bibinfo{person}{Jamy Li} {and} \bibinfo{person}{Mark
  Chignell}.} \bibinfo{year}{2022}\natexlab{}.
\newblock \showarticletitle{{FMEA-AI}: {AI} fairness impact assessment using
  failure mode and effects analysis}.
\newblock \bibinfo{journal}{\emph{AI and Ethics}} (\bibinfo{date}{March}
  \bibinfo{year}{2022}).
\newblock


\bibitem[Liang et~al\mbox{.}(2020)]%
        {Liang2020-xf}
\bibfield{author}{\bibinfo{person}{Paul~Pu Liang},
  \bibinfo{person}{Irene~Mengze Li}, \bibinfo{person}{Emily Zheng},
  \bibinfo{person}{Yao~Chong Lim}, \bibinfo{person}{Ruslan Salakhutdinov},
  {and} \bibinfo{person}{Louis-Philippe Morency}.}
  \bibinfo{year}{2020}\natexlab{}.
\newblock \showarticletitle{Towards Debiasing Sentence Representations}.
\newblock  (\bibinfo{date}{July} \bibinfo{year}{2020}).
\newblock
\showeprint[arxiv]{2007.08100}~[cs.CL]


\bibitem[Liao et~al\mbox{.}(2020)]%
        {Liao2020-cy}
\bibfield{author}{\bibinfo{person}{Q~Vera Liao}, \bibinfo{person}{Daniel
  Gruen}, {and} \bibinfo{person}{Sarah Miller}.}
  \bibinfo{year}{2020}\natexlab{}.
\newblock \showarticletitle{Questioning the {AI}: Informing Design Practices
  for Explainable {AI} User Experiences}. In
  \bibinfo{booktitle}{\emph{Proceedings of the 2020 {CHI} Conference on Human
  Factors in Computing Systems}} (Honolulu, HI, USA)
  \emph{(\bibinfo{series}{CHI '20})}. \bibinfo{publisher}{Association for
  Computing Machinery}, \bibinfo{address}{New York, NY, USA},
  \bibinfo{pages}{1--15}.
\newblock


\bibitem[Liao et~al\mbox{.}(2021)]%
        {liao2021we}
\bibfield{author}{\bibinfo{person}{Thomas Liao}, \bibinfo{person}{Rohan Taori},
  \bibinfo{person}{Inioluwa~Deborah Raji}, {and} \bibinfo{person}{Ludwig
  Schmidt}.} \bibinfo{year}{2021}\natexlab{}.
\newblock \showarticletitle{Are we learning yet? a meta review of evaluation
  failures across machine learning}. In \bibinfo{booktitle}{\emph{Thirty-fifth
  Conference on Neural Information Processing Systems Datasets and Benchmarks
  Track (Round 2)}}.
\newblock


\bibitem[Lo et~al\mbox{.}(2021)]%
        {Hindawi_undated-gp}
\bibfield{author}{\bibinfo{person}{Huai-Wei Lo}, \bibinfo{person}{James J~H
  Liou}, \bibinfo{person}{Jen-Jen Yang}, \bibinfo{person}{Chun-Nen Huang},
  {and} \bibinfo{person}{Yu-Hsuan Lu}.} \bibinfo{year}{2021}\natexlab{}.
\newblock \showarticletitle{An Extended FMEA Model for Exploring the Potential
  Failure Modes: A Case Study of a Steam Turbine for a Nuclear Power Plant}.
\newblock \bibinfo{journal}{\emph{Hindawi}} (\bibinfo{year}{2021}).
\newblock


\bibitem[Madaio et~al\mbox{.}(2022)]%
        {Madaio2022-yz}
\bibfield{author}{\bibinfo{person}{Michael Madaio}, \bibinfo{person}{Lisa
  Egede}, \bibinfo{person}{Hariharan Subramonyam},
  \bibinfo{person}{Jennifer~Wortman Vaughan}, {and} \bibinfo{person}{Hanna
  Wallach}.} \bibinfo{year}{2022}\natexlab{}.
\newblock \bibinfo{title}{Assessing the Fairness of {AI} Systems: {AI}
  Practitioners' Processes, Challenges, and Needs for Support}.
\newblock , \bibinfo{numpages}{26}~pages.
\newblock


\bibitem[Madaio et~al\mbox{.}(2020)]%
        {Madaio2020}
\bibfield{author}{\bibinfo{person}{Michael~A. Madaio}, \bibinfo{person}{Luke
  Stark}, \bibinfo{person}{Jennifer Wortman~Vaughan}, {and}
  \bibinfo{person}{Hanna Wallach}.} \bibinfo{year}{2020}\natexlab{}.
\newblock \showarticletitle{Co-Designing Checklists to Understand
  Organizational Challenges and Opportunities around Fairness in AI}. In
  \bibinfo{booktitle}{\emph{Proceedings of the 2020 CHI Conference on Human
  Factors in Computing Systems}} (Honolulu, HI, USA)
  \emph{(\bibinfo{series}{CHI '20})}. \bibinfo{publisher}{Association for
  Computing Machinery}, \bibinfo{address}{New York, NY, USA},
  \bibinfo{pages}{1–14}.
\newblock
\showISBNx{9781450367080}
\urldef\tempurl%
\url{https://doi.org/10.1145/3313831.3376445}
\showDOI{\tempurl}


\bibitem[Mann and Matzner(2019)]%
        {Mann2019-ns}
\bibfield{author}{\bibinfo{person}{Monique Mann} {and} \bibinfo{person}{Tobias
  Matzner}.} \bibinfo{year}{2019}\natexlab{}.
\newblock \showarticletitle{Challenging algorithmic profiling: The limits of
  data protection and anti-discrimination in responding to emergent
  discrimination}.
\newblock \bibinfo{journal}{\emph{Big Data \& Society}} \bibinfo{volume}{6},
  \bibinfo{number}{2} (\bibinfo{date}{July} \bibinfo{year}{2019}),
  \bibinfo{pages}{2053951719895805}.
\newblock


\bibitem[Mantelero(2022)]%
        {Mantelero2022-en}
\bibfield{author}{\bibinfo{person}{Alessandro Mantelero}.}
  \bibinfo{year}{2022}\natexlab{}.
\newblock \showarticletitle{Human Rights Impact Assessment and {AI}}.
\newblock In \bibinfo{booktitle}{\emph{Beyond Data: Human Rights, Ethical and
  Social Impact Assessment in {AI}}},
  \bibfield{editor}{\bibinfo{person}{Alessandro Mantelero}} (Ed.).
  \bibinfo{publisher}{T.M.C. Asser Press}, \bibinfo{address}{The Hague},
  \bibinfo{pages}{45--91}.
\newblock


\bibitem[Martelaro et~al\mbox{.}(2022)]%
        {Martelaro2022}
\bibfield{author}{\bibinfo{person}{Nikolas Martelaro},
  \bibinfo{person}{Carol~J. Smith}, {and} \bibinfo{person}{Tamara Zilovic}.}
  \bibinfo{year}{2022}\natexlab{}.
\newblock \bibinfo{title}{Exploring Opportunities in Usable Hazard Analysis
  Processes for AI Engineering}.
\newblock
\newblock
\urldef\tempurl%
\url{https://doi.org/10.48550/ARXIV.2203.15628}
\showDOI{\tempurl}


\bibitem[Martin et~al\mbox{.}(2020)]%
        {Martin2020-dl}
\bibfield{author}{\bibinfo{person}{Donald Martin, Jr},
  \bibinfo{person}{Vinodkumar Prabhakaran}, \bibinfo{person}{Jill Kuhlberg},
  \bibinfo{person}{Andrew Smart}, {and} \bibinfo{person}{William~S Isaac}.}
  \bibinfo{year}{2020}\natexlab{}.
\newblock \showarticletitle{Participatory Problem Formulation for Fairer
  Machine Learning Through Community Based System Dynamics}.
\newblock  (\bibinfo{date}{May} \bibinfo{year}{2020}).
\newblock
\showeprint[arxiv]{2005.07572}~[cs.CY]


\bibitem[Metcalf et~al\mbox{.}(2019)]%
        {Metcalf2019-eo}
\bibfield{author}{\bibinfo{person}{Jacob Metcalf}, \bibinfo{person}{Emanuel
  Moss}, {and} \bibinfo{person}{Danah Boyd}.} \bibinfo{year}{2019}\natexlab{}.
\newblock \showarticletitle{Owning Ethics: Corporate Logics, Silicon Valley,
  and the Institutionalization of Ethics}.
\newblock \bibinfo{journal}{\emph{Social Research: An International Quarterly}}
  \bibinfo{volume}{86}, \bibinfo{number}{2} (\bibinfo{year}{2019}),
  \bibinfo{pages}{449--476}.
\newblock


\bibitem[Mitchell et~al\mbox{.}(2019)]%
        {Mitchell2019-ms}
\bibfield{author}{\bibinfo{person}{Margaret Mitchell}, \bibinfo{person}{Simone
  Wu}, \bibinfo{person}{Andrew Zaldivar}, \bibinfo{person}{Parker Barnes},
  \bibinfo{person}{Lucy Vasserman}, \bibinfo{person}{Ben Hutchinson},
  \bibinfo{person}{Elena Spitzer}, \bibinfo{person}{Inioluwa~Deborah Raji},
  {and} \bibinfo{person}{Timnit Gebru}.} \bibinfo{year}{2019}\natexlab{}.
\newblock \showarticletitle{Model Cards for Model Reporting}. In
  \bibinfo{booktitle}{\emph{Proceedings of the Conference on Fairness,
  Accountability, and Transparency}} (Atlanta, GA, USA)
  \emph{(\bibinfo{series}{FAT* '19})}. \bibinfo{publisher}{Association for
  Computing Machinery}, \bibinfo{address}{New York, NY, USA},
  \bibinfo{pages}{220--229}.
\newblock


\bibitem[Mohamed et~al\mbox{.}(2020)]%
        {Mohamed2020-uu}
\bibfield{author}{\bibinfo{person}{Shakir Mohamed},
  \bibinfo{person}{Marie-Therese Png}, {and} \bibinfo{person}{William Isaac}.}
  \bibinfo{year}{2020}\natexlab{}.
\newblock \showarticletitle{Decolonial {AI}: Decolonial Theory as
  Sociotechnical Foresight in Artificial Intelligence}.
\newblock \bibinfo{journal}{\emph{Philos. Technol.}} \bibinfo{volume}{33},
  \bibinfo{number}{4} (\bibinfo{date}{Dec.} \bibinfo{year}{2020}),
  \bibinfo{pages}{659--684}.
\newblock


\bibitem[Molnar et~al\mbox{.}(2020)]%
        {Molnar2020-rb}
\bibfield{author}{\bibinfo{person}{Christoph Molnar}, \bibinfo{person}{Giuseppe
  Casalicchio}, {and} \bibinfo{person}{Bernd Bischl}.}
  \bibinfo{year}{2020}\natexlab{}.
\newblock \showarticletitle{Quantifying Model Complexity via Functional
  Decomposition for Better Post-hoc Interpretability}. In
  \bibinfo{booktitle}{\emph{Machine Learning and Knowledge Discovery in
  Databases}}. \bibinfo{publisher}{Springer International Publishing},
  \bibinfo{pages}{193--204}.
\newblock


\bibitem[Moss and Metcalf(2020)]%
        {Moss2020-le}
\bibfield{author}{\bibinfo{person}{Emanuel Moss} {and} \bibinfo{person}{Jacob
  Metcalf}.} \bibinfo{year}{2020}\natexlab{}.
\newblock \bibinfo{booktitle}{\emph{Ethics owners: a new model of
  organizational responsibility in data-driven technology companies}}.
\newblock \bibinfo{type}{{T}echnical {R}eport}. \bibinfo{institution}{Data \&
  Society Research Institute}.
\newblock


\bibitem[Moss et~al\mbox{.}(2021)]%
        {Moss2021-qi}
\bibfield{author}{\bibinfo{person}{Emanuel Moss},
  \bibinfo{person}{Elizabeth~Anne Watkins}, \bibinfo{person}{Ranjit Singh},
  \bibinfo{person}{Madeleine~Clare Elish}, {and} \bibinfo{person}{Jacob
  Metcalf}.} \bibinfo{year}{2021}\natexlab{}.
\newblock \bibinfo{title}{Assembling Accountability: Algorithmic Impact
  Assessment for the Public Interest}.  (\bibinfo{date}{June}
  \bibinfo{year}{2021}).
\newblock


\bibitem[Nahar et~al\mbox{.}(2022)]%
        {Nahar2022}
\bibfield{author}{\bibinfo{person}{Nadia Nahar}, \bibinfo{person}{Shurui Zhou},
  \bibinfo{person}{Grace Lewis}, {and} \bibinfo{person}{Christian
  K\"{a}stner}.} \bibinfo{year}{2022}\natexlab{}.
\newblock \showarticletitle{Collaboration Challenges in Building ML-Enabled
  Systems: Communication, Documentation, Engineering, and Process}. In
  \bibinfo{booktitle}{\emph{Proceedings of the 44th International Conference on
  Software Engineering}} (Pittsburgh, Pennsylvania)
  \emph{(\bibinfo{series}{ICSE '22})}. \bibinfo{publisher}{Association for
  Computing Machinery}, \bibinfo{address}{New York, NY, USA},
  \bibinfo{pages}{413–425}.
\newblock
\showISBNx{9781450392211}
\urldef\tempurl%
\url{https://doi.org/10.1145/3510003.3510209}
\showDOI{\tempurl}


\bibitem[Ngo et~al\mbox{.}(2021)]%
        {Ngo2021-jj}
\bibfield{author}{\bibinfo{person}{Helen Ngo}, \bibinfo{person}{Cooper
  Raterink}, \bibinfo{person}{Jo{\~a}o G~M Ara{\'u}jo}, \bibinfo{person}{Ivan
  Zhang}, \bibinfo{person}{Carol Chen}, \bibinfo{person}{Adrien Morisot}, {and}
  \bibinfo{person}{Nicholas Frosst}.} \bibinfo{year}{2021}\natexlab{}.
\newblock \showarticletitle{Mitigating harm in language models with
  conditional-likelihood filtration}.
\newblock  (\bibinfo{date}{Aug.} \bibinfo{year}{2021}).
\newblock
\showeprint[arxiv]{2108.07790}~[cs.CL]


\bibitem[Nissenbaum(2001)]%
        {Nissenbaum2001-dm}
\bibfield{author}{\bibinfo{person}{H Nissenbaum}.}
  \bibinfo{year}{2001}\natexlab{}.
\newblock \showarticletitle{How computer systems embody values}.
\newblock \bibinfo{journal}{\emph{Computer}} \bibinfo{volume}{34},
  \bibinfo{number}{3} (\bibinfo{date}{March} \bibinfo{year}{2001}),
  \bibinfo{pages}{120--119}.
\newblock


\bibitem[Noble(2018)]%
        {Noble2018-xs}
\bibfield{author}{\bibinfo{person}{Safiya~Umoja Noble}.}
  \bibinfo{year}{2018}\natexlab{}.
\newblock \bibinfo{booktitle}{\emph{Algorithms of Oppression: How Search
  Engines Reinforce Racism}}.
\newblock \bibinfo{publisher}{NYU Press}.
\newblock


\bibitem[Orlikowski(2000)]%
        {Orlikowski2000-ic}
\bibfield{author}{\bibinfo{person}{Wanda~J Orlikowski}.}
  \bibinfo{year}{2000}\natexlab{}.
\newblock \showarticletitle{Using Technology and Constituting Structures: A
  Practice Lens for Studying Technology in Organizations}.
\newblock \bibinfo{journal}{\emph{Organization Science}} \bibinfo{volume}{11},
  \bibinfo{number}{4} (\bibinfo{year}{2000}), \bibinfo{pages}{404--428}.
\newblock


\bibitem[Patriarca et~al\mbox{.}(2022)]%
        {Patriarca2022-mv}
\bibfield{author}{\bibinfo{person}{Riccardo Patriarca}, \bibinfo{person}{Mikela
  Chatzimichailidou}, \bibinfo{person}{Nektarios Karanikas}, {and}
  \bibinfo{person}{Giulio Di~Gravio}.} \bibinfo{year}{2022}\natexlab{}.
\newblock \showarticletitle{The past and present of {System-Theoretic} Accident
  Model And Processes ({STAMP}) and its associated techniques: A scoping
  review}.
\newblock \bibinfo{journal}{\emph{Saf. Sci.}}  \bibinfo{volume}{146}
  (\bibinfo{date}{Feb.} \bibinfo{year}{2022}), \bibinfo{pages}{105566}.
\newblock


\bibitem[Paullada et~al\mbox{.}(2021)]%
        {Paullada2021-eu}
\bibfield{author}{\bibinfo{person}{Amandalynne Paullada},
  \bibinfo{person}{Inioluwa~Deborah Raji}, \bibinfo{person}{Emily~M Bender},
  \bibinfo{person}{Emily Denton}, {and} \bibinfo{person}{Alex Hanna}.}
  \bibinfo{year}{2021}\natexlab{}.
\newblock \showarticletitle{Data and its (dis)contents: A survey of dataset
  development and use in machine learning research}.
\newblock \bibinfo{journal}{\emph{Patterns (N Y)}} \bibinfo{volume}{2},
  \bibinfo{number}{11} (\bibinfo{date}{Nov.} \bibinfo{year}{2021}),
  \bibinfo{pages}{100336}.
\newblock


\bibitem[Pawlicki et~al\mbox{.}(2016)]%
        {Pawlicki2016-ef}
\bibfield{author}{\bibinfo{person}{Todd Pawlicki}, \bibinfo{person}{Aubrey
  Samost}, \bibinfo{person}{Derek~W Brown}, \bibinfo{person}{Ryan~P Manger},
  \bibinfo{person}{Gwe-Ya Kim}, {and} \bibinfo{person}{Nancy~G Leveson}.}
  \bibinfo{year}{2016}\natexlab{}.
\newblock \showarticletitle{Application of systems and control theory-based
  hazard analysis to radiation oncology}.
\newblock \bibinfo{journal}{\emph{Med. Phys.}} \bibinfo{volume}{43},
  \bibinfo{number}{3} (\bibinfo{date}{March} \bibinfo{year}{2016}),
  \bibinfo{pages}{1514--1530}.
\newblock


\bibitem[Perrow(1984)]%
        {perrow1984normal}
\bibfield{author}{\bibinfo{person}{Charles Perrow}.}
  \bibinfo{year}{1984}\natexlab{}.
\newblock \bibinfo{booktitle}{\emph{Normal accidents: Living with high risk
  technologies}}.
\newblock \bibinfo{publisher}{Basic Books}, \bibinfo{address}{New York}.
\newblock


\bibitem[Poursabzi-Sangdeh et~al\mbox{.}(2018)]%
        {Poursabzi-Sangdeh2018-ks}
\bibfield{author}{\bibinfo{person}{Forough Poursabzi-Sangdeh},
  \bibinfo{person}{Daniel~G Goldstein}, \bibinfo{person}{Jake~M Hofman},
  \bibinfo{person}{Jennifer~Wortman Vaughan}, {and} \bibinfo{person}{Hanna
  Wallach}.} \bibinfo{year}{2018}\natexlab{}.
\newblock \showarticletitle{Manipulating and Measuring Model Interpretability}.
\newblock  (\bibinfo{date}{Feb.} \bibinfo{year}{2018}).
\newblock
\showeprint[arxiv]{1802.07810}~[cs.AI]


\bibitem[Raji et~al\mbox{.}(2022)]%
        {Raji2022-cz}
\bibfield{author}{\bibinfo{person}{Inioluwa~Deborah Raji},
  \bibinfo{person}{I~Elizabeth Kumar}, \bibinfo{person}{Aaron Horowitz}, {and}
  \bibinfo{person}{Andrew Selbst}.} \bibinfo{year}{2022}\natexlab{}.
\newblock \showarticletitle{The Fallacy of {AI} Functionality}. In
  \bibinfo{booktitle}{\emph{2022 {ACM} Conference on Fairness, Accountability,
  and Transparency}} (Seoul, Republic of Korea) \emph{(\bibinfo{series}{FAccT
  '22})}. \bibinfo{publisher}{Association for Computing Machinery},
  \bibinfo{address}{New York, NY, USA}, \bibinfo{pages}{959--972}.
\newblock


\bibitem[Raji et~al\mbox{.}(2020a)]%
        {Raji2020-dw}
\bibfield{author}{\bibinfo{person}{Inioluwa~Deborah Raji},
  \bibinfo{person}{Andrew Smart}, \bibinfo{person}{Rebecca~N White},
  \bibinfo{person}{Margaret Mitchell}, \bibinfo{person}{Timnit Gebru},
  \bibinfo{person}{Ben Hutchinson}, \bibinfo{person}{Jamila Smith-Loud},
  \bibinfo{person}{Daniel Theron}, {and} \bibinfo{person}{Parker Barnes}.}
  \bibinfo{year}{2020}\natexlab{a}.
\newblock \showarticletitle{Closing the {AI} accountability gap: defining an
  end-to-end framework for internal algorithmic auditing}. In
  \bibinfo{booktitle}{\emph{Proceedings of the 2020 Conference on Fairness,
  Accountability, and Transparency}} (Barcelona, Spain)
  \emph{(\bibinfo{series}{FAT* '20})}. \bibinfo{publisher}{Association for
  Computing Machinery}, \bibinfo{address}{New York, NY, USA},
  \bibinfo{pages}{33--44}.
\newblock


\bibitem[Raji et~al\mbox{.}(2020b)]%
        {Raji2020closing}
\bibfield{author}{\bibinfo{person}{Inioluwa~Deborah Raji},
  \bibinfo{person}{Andrew Smart}, \bibinfo{person}{Rebecca~N. White},
  \bibinfo{person}{Margaret Mitchell}, \bibinfo{person}{Timnit Gebru},
  \bibinfo{person}{Ben Hutchinson}, \bibinfo{person}{Jamila Smith-Loud},
  \bibinfo{person}{Daniel Theron}, {and} \bibinfo{person}{Parker Barnes}.}
  \bibinfo{year}{2020}\natexlab{b}.
\newblock \showarticletitle{Closing the AI Accountability Gap: Defining an
  End-to-End Framework for Internal Algorithmic Auditing}. In
  \bibinfo{booktitle}{\emph{Proceedings of the 2020 Conference on Fairness,
  Accountability, and Transparency}} (Barcelona, Spain)
  \emph{(\bibinfo{series}{FAT* '20})}. \bibinfo{publisher}{Association for
  Computing Machinery}, \bibinfo{address}{New York, NY, USA},
  \bibinfo{pages}{33–44}.
\newblock
\showISBNx{9781450369367}
\urldef\tempurl%
\url{https://doi.org/10.1145/3351095.3372873}
\showDOI{\tempurl}


\bibitem[Rakova et~al\mbox{.}(2021a)]%
        {Rakova2021}
\bibfield{author}{\bibinfo{person}{Bogdana Rakova}, \bibinfo{person}{Jingying
  Yang}, \bibinfo{person}{Henriette Cramer}, {and} \bibinfo{person}{Rumman
  Chowdhury}.} \bibinfo{year}{2021}\natexlab{a}.
\newblock \showarticletitle{Where Responsible AI Meets Reality: Practitioner
  Perspectives on Enablers for Shifting Organizational Practices}.
\newblock \bibinfo{journal}{\emph{Proc. ACM Hum.-Comput. Interact.}}
  \bibinfo{volume}{5}, \bibinfo{number}{CSCW1}, Article \bibinfo{articleno}{7}
  (\bibinfo{date}{apr} \bibinfo{year}{2021}), \bibinfo{numpages}{23}~pages.
\newblock
\urldef\tempurl%
\url{https://doi.org/10.1145/3449081}
\showDOI{\tempurl}


\bibitem[Rakova et~al\mbox{.}(2021b)]%
        {Rakova2021-ov}
\bibfield{author}{\bibinfo{person}{Bogdana Rakova}, \bibinfo{person}{Jingying
  Yang}, \bibinfo{person}{Henriette Cramer}, {and} \bibinfo{person}{Rumman
  Chowdhury}.} \bibinfo{year}{2021}\natexlab{b}.
\newblock \showarticletitle{Where Responsible {AI} meets Reality: Practitioner
  Perspectives on Enablers for Shifting Organizational Practices}.
\newblock \bibinfo{journal}{\emph{Proc. ACM Hum.-Comput. Interact.}}
  \bibinfo{volume}{5}, \bibinfo{number}{CSCW1} (\bibinfo{date}{April}
  \bibinfo{year}{2021}), \bibinfo{pages}{1--23}.
\newblock


\bibitem[Reader et~al\mbox{.}(2022)]%
        {reader2022models}
\bibfield{author}{\bibinfo{person}{Lydia Reader}, \bibinfo{person}{Pegah
  Nokhiz}, \bibinfo{person}{Cathleen Power}, \bibinfo{person}{Neal Patwari},
  \bibinfo{person}{Suresh Venkatasubramanian}, {and} \bibinfo{person}{Sorelle
  Friedler}.} \bibinfo{year}{2022}\natexlab{}.
\newblock \showarticletitle{Models for understanding and quantifying feedback
  in societal systems}. In \bibinfo{booktitle}{\emph{2022 ACM Conference on
  Fairness, Accountability, and Transparency}}. \bibinfo{pages}{1765--1775}.
\newblock


\bibitem[Rismani and Moon(2021)]%
        {Rismani2021-qy}
\bibfield{author}{\bibinfo{person}{Shalaleh Rismani} {and}
  \bibinfo{person}{Ajung Moon}.} \bibinfo{year}{2021}\natexlab{}.
\newblock \showarticletitle{How do {AI} systems fail socially?: an engineering
  risk analysis approach}. In \bibinfo{booktitle}{\emph{2021 {IEEE}
  International Symposium on Ethics in Engineering, Science and Technology
  ({ETHICS})}}. \bibinfo{pages}{1--8}.
\newblock


\bibitem[Rodrigues et~al\mbox{.}(2012)]%
        {rodrigues2012commercial}
\bibfield{author}{\bibinfo{person}{Clarence~C Rodrigues},
  \bibinfo{person}{Stephen~K Cusick}, {et~al\mbox{.}}}
  \bibinfo{year}{2012}\natexlab{}.
\newblock \bibinfo{booktitle}{\emph{Commercial aviation safety}}.
\newblock \bibinfo{publisher}{McGraw-Hill Education}.
\newblock


\bibitem[Rostamzadeh et~al\mbox{.}(2021)]%
        {rostamzadeh2021thinking}
\bibfield{author}{\bibinfo{person}{Negar Rostamzadeh}, \bibinfo{person}{Ben
  Hutchinson}, \bibinfo{person}{Christina Greer}, {and}
  \bibinfo{person}{Vinodkumar Prabhakaran}.} \bibinfo{year}{2021}\natexlab{}.
\newblock \showarticletitle{Thinking Beyond Distributions in Testing Machine
  Learned Models}.
\newblock \bibinfo{journal}{\emph{arXiv preprint arXiv:2112.03057}}
  (\bibinfo{year}{2021}).
\newblock


\bibitem[Ruiz et~al\mbox{.}(2022)]%
        {ruiz2022simulated}
\bibfield{author}{\bibinfo{person}{Nataniel Ruiz}, \bibinfo{person}{Adam
  Kortylewski}, \bibinfo{person}{Weichao Qiu}, \bibinfo{person}{Cihang Xie},
  \bibinfo{person}{Sarah~Adel Bargal}, \bibinfo{person}{Alan Yuille}, {and}
  \bibinfo{person}{Stan Sclaroff}.} \bibinfo{year}{2022}\natexlab{}.
\newblock \showarticletitle{Simulated Adversarial Testing of Face Recognition
  Models}.
\newblock \bibinfo{journal}{\emph{CVPR}} (\bibinfo{year}{2022}).
\newblock


\bibitem[Sambasivan et~al\mbox{.}(2021a)]%
        {Sambasivan2021-gm}
\bibfield{author}{\bibinfo{person}{Nithya Sambasivan}, \bibinfo{person}{Erin
  Arnesen}, \bibinfo{person}{Ben Hutchinson}, \bibinfo{person}{Tulsee Doshi},
  {and} \bibinfo{person}{Vinodkumar Prabhakaran}.}
  \bibinfo{year}{2021}\natexlab{a}.
\newblock \showarticletitle{Re-imagining Algorithmic Fairness in India and
  Beyond}. In \bibinfo{booktitle}{\emph{Proceedings of the 2021 {ACM}
  Conference on Fairness, Accountability, and Transparency}} (Virtual Event,
  Canada) \emph{(\bibinfo{series}{FAccT '21})}. \bibinfo{publisher}{Association
  for Computing Machinery}, \bibinfo{address}{New York, NY, USA},
  \bibinfo{pages}{315--328}.
\newblock


\bibitem[Sambasivan et~al\mbox{.}(2021b)]%
        {sambasivan2021re}
\bibfield{author}{\bibinfo{person}{Nithya Sambasivan}, \bibinfo{person}{Erin
  Arnesen}, \bibinfo{person}{Ben Hutchinson}, \bibinfo{person}{Tulsee Doshi},
  {and} \bibinfo{person}{Vinodkumar Prabhakaran}.}
  \bibinfo{year}{2021}\natexlab{b}.
\newblock \showarticletitle{Re-Imagining Algorithmic Fairness in India and
  Beyond}. In \bibinfo{booktitle}{\emph{Proceedings of the 2021 ACM Conference
  on Fairness, Accountability, and Transparency}} (Virtual Event, Canada)
  \emph{(\bibinfo{series}{FAccT '21})}. \bibinfo{publisher}{Association for
  Computing Machinery}, \bibinfo{address}{New York, NY, USA},
  \bibinfo{pages}{315–328}.
\newblock
\showISBNx{9781450383097}
\urldef\tempurl%
\url{https://doi.org/10.1145/3442188.3445896}
\showDOI{\tempurl}


\bibitem[Selbst(2020)]%
        {selbst2020negligence}
\bibfield{author}{\bibinfo{person}{Andrew~D Selbst}.}
  \bibinfo{year}{2020}\natexlab{}.
\newblock \showarticletitle{Negligence and AI's human users}.
\newblock \bibinfo{journal}{\emph{BUL Rev.}}  \bibinfo{volume}{100}
  (\bibinfo{year}{2020}), \bibinfo{pages}{1315}.
\newblock


\bibitem[Selbst et~al\mbox{.}(2019)]%
        {Selbst2019-dr}
\bibfield{author}{\bibinfo{person}{Andrew~D Selbst}, \bibinfo{person}{Danah
  Boyd}, \bibinfo{person}{Sorelle~A Friedler}, \bibinfo{person}{Suresh
  Venkatasubramanian}, {and} \bibinfo{person}{Janet Vertesi}.}
  \bibinfo{year}{2019}\natexlab{}.
\newblock \showarticletitle{Fairness and Abstraction in Sociotechnical
  Systems}. In \bibinfo{booktitle}{\emph{Proceedings of the Conference on
  Fairness, Accountability, and Transparency}} (Atlanta, GA, USA)
  \emph{(\bibinfo{series}{FAT* '19})}. \bibinfo{publisher}{Association for
  Computing Machinery}, \bibinfo{address}{New York, NY, USA},
  \bibinfo{pages}{59--68}.
\newblock


\bibitem[Shen et~al\mbox{.}(2021)]%
        {Shen2021-us}
\bibfield{author}{\bibinfo{person}{Hong Shen}, \bibinfo{person}{Alicia DeVos},
  \bibinfo{person}{Motahhare Eslami}, {and} \bibinfo{person}{Kenneth
  Holstein}.} \bibinfo{year}{2021}\natexlab{}.
\newblock \showarticletitle{Everyday Algorithm Auditing: Understanding the
  Power of Everyday Users in Surfacing Harmful Algorithmic Behaviors}.
\newblock \bibinfo{journal}{\emph{Proc. ACM Hum.-Comput. Interact.}}
  \bibinfo{volume}{5}, \bibinfo{number}{CSCW2} (\bibinfo{date}{Oct.}
  \bibinfo{year}{2021}), \bibinfo{pages}{1--29}.
\newblock


\bibitem[Shilton(2013)]%
        {Shilton2013-ez}
\bibfield{author}{\bibinfo{person}{Katie Shilton}.}
  \bibinfo{year}{2013}\natexlab{}.
\newblock \showarticletitle{Values Levers: Building Ethics into Design}.
\newblock \bibinfo{journal}{\emph{Sci. Technol. Human Values}}
  \bibinfo{volume}{38}, \bibinfo{number}{3} (\bibinfo{date}{May}
  \bibinfo{year}{2013}), \bibinfo{pages}{374--397}.
\newblock


\bibitem[Shin et~al\mbox{.}(2021)]%
        {Shin2021-tg}
\bibfield{author}{\bibinfo{person}{Sung-Min Shin}, \bibinfo{person}{Sang~Hun
  Lee}, \bibinfo{person}{Seung K~I Shin}, \bibinfo{person}{Inseok Jang}, {and}
  \bibinfo{person}{Jinkyun Park}.} \bibinfo{year}{2021}\natexlab{}.
\newblock \showarticletitle{{STPA-Based} Hazard and Importance Analysis on
  {NPP} Safety {I\&C} Systems Focusing on {Human--System} Interactions}.
\newblock \bibinfo{journal}{\emph{Reliab. Eng. Syst. Saf.}}
  \bibinfo{volume}{213} (\bibinfo{date}{Sept.} \bibinfo{year}{2021}),
  \bibinfo{pages}{107698}.
\newblock


\bibitem[Shrader-Frechette(1991)]%
        {shrader1991risk}
\bibfield{author}{\bibinfo{person}{Kristin~Sharon Shrader-Frechette}.}
  \bibinfo{year}{1991}\natexlab{}.
\newblock \bibinfo{booktitle}{\emph{Risk and rationality: Philosophical
  foundations for populist reforms}}.
\newblock \bibinfo{publisher}{Univ of California Press}.
\newblock


\bibitem[Spiel et~al\mbox{.}(2019)]%
        {Spiel2019-jk}
\bibfield{author}{\bibinfo{person}{Katta Spiel}, \bibinfo{person}{Alex Ahmed},
  \bibinfo{person}{Jennifer Rode}, {and} \bibinfo{person}{Jean Hardy}.}
  \bibinfo{year}{2019}\natexlab{}.
\newblock \showarticletitle{Queer(ing) {HCI}: Moving Forward in Theory and
  Practice}. In \bibinfo{booktitle}{\emph{Extended Abstracts of the 2019 {CHI}
  Conference}}. \bibinfo{publisher}{unknown}, \bibinfo{pages}{1--4}.
\newblock


\bibitem[Styhre(2018)]%
        {styhre2018unfinished}
\bibfield{author}{\bibinfo{person}{Alexander Styhre}.}
  \bibinfo{year}{2018}\natexlab{}.
\newblock \bibinfo{booktitle}{\emph{The Unfinished Business of Governance:
  Monitoring and Regulating Industries and Organizations}}.
\newblock \bibinfo{publisher}{Edward Elgar Publishing}.
\newblock


\bibitem[Sugiyama et~al\mbox{.}(2007)]%
        {sugiyama2007direct}
\bibfield{author}{\bibinfo{person}{Masashi Sugiyama}, \bibinfo{person}{Shinichi
  Nakajima}, \bibinfo{person}{Hisashi Kashima}, \bibinfo{person}{Paul Buenau},
  {and} \bibinfo{person}{Motoaki Kawanabe}.} \bibinfo{year}{2007}\natexlab{}.
\newblock \showarticletitle{Direct Importance Estimation with Model Selection
  and Its Application to Covariate Shift Adaptation}.
\newblock \bibinfo{journal}{\emph{Advances in Neural Information Processing
  Systems}}  \bibinfo{volume}{20} (\bibinfo{year}{2007}).
\newblock


\bibitem[Sulaman et~al\mbox{.}(2019)]%
        {Sulaman2019-rm}
\bibfield{author}{\bibinfo{person}{Sardar~Muhammad Sulaman},
  \bibinfo{person}{Armin Beer}, \bibinfo{person}{Michael Felderer}, {and}
  \bibinfo{person}{Martin H{\"o}st}.} \bibinfo{year}{2019}\natexlab{}.
\newblock \showarticletitle{Comparison of the {FMEA} and {STPA} safety analysis
  methods--a case study}.
\newblock \bibinfo{journal}{\emph{Software Quality Journal}}
  \bibinfo{volume}{27}, \bibinfo{number}{1} (\bibinfo{date}{March}
  \bibinfo{year}{2019}), \bibinfo{pages}{349--387}.
\newblock


\bibitem[Suresh and Guttag(2021)]%
        {Suresh2021-js}
\bibfield{author}{\bibinfo{person}{Harini Suresh} {and} \bibinfo{person}{John
  Guttag}.} \bibinfo{year}{2021}\natexlab{}.
\newblock \showarticletitle{A Framework for Understanding Sources of Harm
  throughout the Machine Learning Life Cycle}. In
  \bibinfo{booktitle}{\emph{Equity and Access in Algorithms, Mechanisms, and
  Optimization}} (--, NY, USA) \emph{(\bibinfo{series}{EAAMO '21},
  \bibinfo{number}{Article 17})}. \bibinfo{publisher}{Association for Computing
  Machinery}, \bibinfo{address}{New York, NY, USA}, \bibinfo{pages}{1--9}.
\newblock


\bibitem[Tatman(2017)]%
        {Tatman2017-eg}
\bibfield{author}{\bibinfo{person}{Rachael Tatman}.}
  \bibinfo{year}{2017}\natexlab{}.
\newblock \showarticletitle{Gender and Dialect Bias in {YouTube's} Automatic
  Captions}. In \bibinfo{booktitle}{\emph{Proceedings of the First {ACL}
  Workshop on Ethics in Natural Language Processing}}.
  \bibinfo{publisher}{Association for Computational Linguistics},
  \bibinfo{address}{Valencia, Spain}, \bibinfo{pages}{53--59}.
\newblock


\bibitem[{Treasury Board of Canada Secretariat}(2019)]%
        {Treasury_Board_of_Canada_Secretariat2019-sk}
\bibfield{author}{\bibinfo{person}{{Treasury Board of Canada Secretariat}}.}
  \bibinfo{year}{2019}\natexlab{}.
\newblock \bibinfo{title}{Directive on Automated {Decision-Making}}.
\newblock
  \bibinfo{howpublished}{\url{https://www.tbs-sct.canada.ca/pol/doc-eng.aspx?id=32592}}.
\newblock
\newblock
\shownote{Accessed: 2022-9-15}.


\bibitem[Umbrello(2019)]%
        {Umbrello2019-vn}
\bibfield{author}{\bibinfo{person}{Steven Umbrello}.}
  \bibinfo{year}{2019}\natexlab{}.
\newblock \showarticletitle{Beneficial Artificial Intelligence Coordination by
  Means of a Value Sensitive Design Approach}.
\newblock \bibinfo{journal}{\emph{Big Data and Cognitive Computing}}
  \bibinfo{volume}{3}, \bibinfo{number}{1} (\bibinfo{date}{Jan.}
  \bibinfo{year}{2019}), \bibinfo{pages}{5}.
\newblock


\bibitem[Vaughan(1996)]%
        {vaughan1996challenger}
\bibfield{author}{\bibinfo{person}{Diane Vaughan}.}
  \bibinfo{year}{1996}\natexlab{}.
\newblock \bibinfo{booktitle}{\emph{The Challenger launch decision: Risky
  technology, culture, and deviance at NASA}}.
\newblock \bibinfo{publisher}{University of Chicago press}.
\newblock


\bibitem[Vaughan(2020)]%
        {Vaughan2020-zs}
\bibfield{author}{\bibinfo{person}{Jenn~Wortman Vaughan}.}
  \bibinfo{year}{2020}\natexlab{}.
\newblock \bibinfo{title}{Transparency and Intelligibility Throughout the
  Machine Learning Life Cycle}.
\newblock
\newblock


\bibitem[Wambsganss et~al\mbox{.}(2021)]%
        {Wambsganss2021}
\bibfield{author}{\bibinfo{person}{Thiemo Wambsganss}, \bibinfo{person}{Anne
  H{\"o}ch}, \bibinfo{person}{Naim Zierau}, {and} \bibinfo{person}{Matthias
  S{\"o}llner}.} \bibinfo{year}{2021}\natexlab{}.
\newblock \showarticletitle{Ethical Design of Conversational Agents: Towards
  Principles for a Value-Sensitive Design}. In
  \bibinfo{booktitle}{\emph{Innovation Through Information Systems}},
  \bibfield{editor}{\bibinfo{person}{Frederik Ahlemann},
  \bibinfo{person}{Reinhard Sch{\"u}tte}, {and} \bibinfo{person}{Stefan
  Stieglitz}} (Eds.). \bibinfo{publisher}{Springer International Publishing},
  \bibinfo{address}{Cham}, \bibinfo{pages}{539--557}.
\newblock
\showISBNx{978-3-030-86790-4}


\bibitem[Weidinger et~al\mbox{.}(2022)]%
        {Weidinger2022-qv}
\bibfield{author}{\bibinfo{person}{Laura Weidinger}, \bibinfo{person}{Jonathan
  Uesato}, \bibinfo{person}{Maribeth Rauh}, \bibinfo{person}{Conor Griffin},
  \bibinfo{person}{Po-Sen Huang}, \bibinfo{person}{John Mellor},
  \bibinfo{person}{Amelia Glaese}, \bibinfo{person}{Myra Cheng},
  \bibinfo{person}{Borja Balle}, \bibinfo{person}{Atoosa Kasirzadeh},
  \bibinfo{person}{Courtney Biles}, \bibinfo{person}{Sasha Brown},
  \bibinfo{person}{Zac Kenton}, \bibinfo{person}{Will Hawkins},
  \bibinfo{person}{Tom Stepleton}, \bibinfo{person}{Abeba Birhane},
  \bibinfo{person}{Lisa~Anne Hendricks}, \bibinfo{person}{Laura Rimell},
  \bibinfo{person}{William Isaac}, \bibinfo{person}{Julia Haas},
  \bibinfo{person}{Sean Legassick}, \bibinfo{person}{Geoffrey Irving}, {and}
  \bibinfo{person}{Iason Gabriel}.} \bibinfo{year}{2022}\natexlab{}.
\newblock \showarticletitle{Taxonomy of Risks posed by Language Models}. In
  \bibinfo{booktitle}{\emph{2022 {ACM} Conference on Fairness, Accountability,
  and Transparency}} (Seoul, Republic of Korea) \emph{(\bibinfo{series}{FAccT
  '22})}. \bibinfo{publisher}{Association for Computing Machinery},
  \bibinfo{address}{New York, NY, USA}, \bibinfo{pages}{214--229}.
\newblock


\bibitem[Wong et~al\mbox{.}(2022)]%
        {Wong2022}
\bibfield{author}{\bibinfo{person}{Richmond~Y. Wong},
  \bibinfo{person}{Michael~A. Madaio}, {and} \bibinfo{person}{Nick Merrill}.}
  \bibinfo{year}{2022}\natexlab{}.
\newblock \bibinfo{title}{Seeing Like a Toolkit: How Toolkits Envision the Work
  of AI Ethics}.
\newblock
\newblock
\urldef\tempurl%
\url{https://doi.org/10.48550/ARXIV.2202.08792}
\showDOI{\tempurl}


\bibitem[Yee et~al\mbox{.}(2021)]%
        {Yee2021-eg}
\bibfield{author}{\bibinfo{person}{Kyra Yee}, \bibinfo{person}{Uthaipon
  Tantipongpipat}, {and} \bibinfo{person}{Shubhanshu Mishra}.}
  \bibinfo{year}{2021}\natexlab{}.
\newblock \showarticletitle{Image Cropping on Twitter: Fairness Metrics, their
  Limitations, and the Importance of Representation, Design, and Agency}.
\newblock  (\bibinfo{date}{May} \bibinfo{year}{2021}).
\newblock
\showeprint[arxiv]{2105.08667}~[cs.CY]


\bibitem[Zeng et~al\mbox{.}(2021)]%
        {zeng2021openattack}
\bibfield{author}{\bibinfo{person}{Guoyang Zeng}, \bibinfo{person}{Fanchao Qi},
  \bibinfo{person}{Qianrui Zhou}, \bibinfo{person}{Tingji Zhang},
  \bibinfo{person}{Zixian Ma}, \bibinfo{person}{Bairu Hou},
  \bibinfo{person}{Yuan Zang}, \bibinfo{person}{Zhiyuan Liu}, {and}
  \bibinfo{person}{Maosong Sun}.} \bibinfo{year}{2021}\natexlab{}.
\newblock \showarticletitle{{OpenAttack}: An Open-source Textual Adversarial
  Attack Toolkit}. In \bibinfo{booktitle}{\emph{Proceedings of the 59th Annual
  Meeting of the Association for Computational Linguistics and the 11th
  International Joint Conference on Natural Language Processing: System
  Demonstrations}}. \bibinfo{pages}{363--371}.
\newblock


\bibitem[Zhang et~al\mbox{.}(2020)]%
        {zhang2020adversarial}
\bibfield{author}{\bibinfo{person}{Wei~Emma Zhang}, \bibinfo{person}{Quan~Z
  Sheng}, \bibinfo{person}{Ahoud Alhazmi}, {and} \bibinfo{person}{Chenliang
  Li}.} \bibinfo{year}{2020}\natexlab{}.
\newblock \showarticletitle{Adversarial attacks on deep-learning models in
  natural language processing: A survey}.
\newblock \bibinfo{journal}{\emph{ACM Transactions on Intelligent Systems and
  Technology (TIST)}} \bibinfo{volume}{11}, \bibinfo{number}{3}
  (\bibinfo{year}{2020}), \bibinfo{pages}{1--41}.
\newblock


\bibitem[Zytko et~al\mbox{.}(2022)]%
        {Zytko2022}
\bibfield{author}{\bibinfo{person}{Douglas Zytko}, \bibinfo{person}{Pamela
  J.~Wisniewski}, \bibinfo{person}{Shion Guha}, \bibinfo{person}{Eric
  P.~S.~Baumer}, {and} \bibinfo{person}{Min~Kyung Lee}.}
  \bibinfo{year}{2022}\natexlab{}.
\newblock \showarticletitle{Participatory Design of AI Systems: Opportunities
  and Challenges Across Diverse Users, Relationships, and Application Domains}.
  In \bibinfo{booktitle}{\emph{Extended Abstracts of the 2022 CHI Conference on
  Human Factors in Computing Systems}} (New Orleans, LA, USA)
  \emph{(\bibinfo{series}{CHI EA '22})}. \bibinfo{publisher}{Association for
  Computing Machinery}, \bibinfo{address}{New York, NY, USA}, Article
  \bibinfo{articleno}{154}, \bibinfo{numpages}{4}~pages.
\newblock
\showISBNx{9781450391566}
\urldef\tempurl%
\url{https://doi.org/10.1145/3491101.3516506}
\showDOI{\tempurl}


\end{thebibliography}

\newpage
\appendix

\section{Study protocol}

\subsection{Interview protocol}
The following outlines the interview protocol and the list of interview questions: 

\begin{quote}

Thank you for taking the time to participate in this interview. 

I am conducting these expert interviews to understand existing social and ethical risk management practices for ML systems and brainstorm on how they could be improved.

I will start by asking you about your current practices in the first part of the interview. 

I will then introduce two risk management tools from reliability and safety engineering. We will spend some time discussing the pros/cons of these techniques and brainstorm about the usability of such techniques within the ML development process. 

\end{quote}
\begin{quote}
    
\textit{Part 1 [15 min] - current practices }

\begin{itemize}
    \item Could you please briefly describe your role and responsibilities in your current position.  
    \item In your work, how do you currently define social and ethical risks for ML systems? How did this definition come about?
    \item What type of ML systems have you assessed in terms of social and ethical risks?What type of ML systems have you worked with?  
    \item How do you currently assess social and ethical risks for ML systems and how were these assessment processes developed? 
    \item How do you currently mitigate social and ethical risks for ML systems and how were these mitigation processes developed? 
    \item What are the challenges that you face when assessing and mitigating ethical and social risks for different projects? How have you dealt with these challenges?

\end{itemize}
\textit{Part 2 [ 30 min] - safety engineering frameworks}

In the rest of this interview we will ask your feedback about two different risk assessment techniques that have mostly been applied within aerospace, medical device and military settings. One of these techniques has a top to bottom approach of examining risks and potential controls while the other one has a bottom up approach. 

In this part, introduce STPA and FMEA using the slide deck shown in section \ref{visualguide}. Introduce each technique one at a time and get feedback on each for 10 minutes using the following questions. 

\begin{itemize}
    \item Have you heard of/ are you familiar with STPA/FMEA? If yes, have you used them? In what context? 
    \item What are your initial impressions of STPA/FMEA and it application for ML systems? 
    \item What are the pros and cons of applying STPA/FMEA for managing ethical and social risk? 
    \item Thinking about an ML application that you have worked with in the past, can you walk me through how you would adapt STPA/FMEA to look at social and ethical risks of this application?

\end{itemize}
\end{quote}
\newpage
\end{document}